\documentclass[onecolumn]{aa} 
\usepackage{graphicx}
\usepackage{epstopdf}
\usepackage{natbib}
\usepackage{txfonts}
\newcommand{\lp}{\left(}
\newcommand{\rp}{\right)}

\newcommand{\vna}{\vec{\nabla}}

\def\vec#1{\ensuremath{\mathchoice{\mbox{\boldmath$\displaystyle#1$}}
{\mbox{\boldmath$\textstyle#1$}}
{\mbox{\boldmath$\scriptstyle#1$}}
{\mbox{\boldmath$\scriptscriptstyle#1$}}}}

\begin{document}
   \title{3D evolution of magnetic fields in a differentially \\
rotating stellar radiative zone}

\author{L. Jouve\inst{\ref{irap1},\ref{irap2}}
\and T. Gastine\inst{\ref{mps}}
\and F. Ligni\`eres\inst{\ref{irap1},\ref{irap2}}}

\institute{Universit\'e de Toulouse, UPS-OMP, Institut de Recherche en Astrophysique et Plan\'etologie, 31028 Toulouse Cedex 4, France\\
email:laurene.jouve@irap.omp.eu\label{irap1}
\and CNRS, Institut de Recherche en Astrophysique et Plan\'etologie, 14 avenue Edouard Belin, 31400 Toulouse, France\label{irap2}
\and Max-Planck-Institut f\"ur Sonnensystemforschung, Justus-von-Liebig-Weg 3, 37077 G\"ottingen, Germany\\
e-mail:gastine@mps.mpg.de\label{mps}}
\date{Received /
Accepted}

\abstract

   \date{Received; accepted }

 
  \abstract
{The question of the origin and evolution of magnetic fields in stars possessing a radiative envelope, like the A-type stars, is still regarded as a challenge for stellar physics.
Those zones are likely to be differentially rotating, which suggests that strong interactions between differential rotation and magnetic fields could be at play in such regions.}
   {We would like to analyze in detail 
the evolution of magnetic fields in a differentially rotating stellar radiative zone and the possible presence of magnetic instabilities.}
  {We numerically compute the joint evolution of the magnetic and velocity 
fields in a 3D spherical shell starting from an initial profile for the poloidal 
magnetic field and differential rotation. 
In order to characterize the nature of the magnetic instabilities
that may be expected, 
we use the predictions of a local linear analysis.}
   {The poloidal magnetic field is initially wound-up by the differential 
rotation to produce a toroidal field which becomes unstable. We find that different types of instabilities may
occur, depending on the balance between the shear strength and the magnetic field intensity.
In the particular setup studied here where the differential rotation is dominant, the instability is not of 
Tayler-type but is the magneto-rotational instability. The growth rate of 
the instability depends mainly on the initial rotation rate, while the background state 
typically oscillates over a poloidal Alfv\'en time. We thus find that 
the axisymmetric magnetic configuration is strongly modified by the instability 
only if the ratio between the poloidal Alfv\'en frequency and the rotation rate is 
sufficiently small. An enhanced transport of angular momentum is found in the 
most unstable cases: the typical time to flatten the rotation profile is 
then much faster than the decay time associated with the phase-mixing mechanism, which also occurs in the stable cases.
 When the instability saturates before reaching a significant amplitude, the magnetic 
configuration relaxes into a stable axisymmetric equilibrium formed by several 
twisted tori.}
   {We conclude that the magneto-rotational instability is always favored (over 
the Tayler instability) in unstratified spherical shells when an initial poloidal field is sheared by 
a sufficiently strong cylindrical differential rotation. A possible application to the magnetic desert observed among A 
stars is given. We argue that the dichotomy between stars exhibiting strong 
axisymmetric fields (Ap stars) and those harboring a sub-Gauss magnetism could 
be linked to the threshold for the instability.}


   \maketitle
%

\section{Introduction}

Dynamical processes associated with the presence of a magnetic field in a stellar radiative zone have been the subject of various studies in the past decades. The main basic questions addressed by these studies are:

- What is the origin of the magnetic field observed at the surface of roughly $10\%$ of intermediate-mass and massive stars which are incapable of producing a convective dynamo as we know it from cooler stars?

- What is the effect of magnetic fields on the transport of angular momentum under the general conditions existing in a stellar radiative zone?

Interests in the first question have recently revived thanks to large 
spectropolarimetric surveys performed with the \emph{Narval} and \emph{ESPaDOnS} 
instruments. The chemically peculiar Ap-Bp stars, representing around $10\%$ of 
the intermediate-mass star population, were confirmed to host invariable 
dipolar-like magnetic fields with intensities ranging from approximately 
$300\rm{G}$ to $30 \rm{kG}$ \citep{Donati09}. Another class of magnetism has now 
been revealed in A-type stars thanks to recent improvements in the detection 
limit of those spectropolarimeters: the so-called Vega-like magnetism, named 
after the bright star in which it was discovered first \citep{Lignieres09, Petit11}. The structure of the 
magnetic field in those stars is fairly different from the strong dipole 
observed at the surface of Ap stars. Vega indeed exhibits a surface magnetic 
field organized at much smaller scales and typical amplitudes for the 
line-of-sight field of less than $1{\rm G}$. In these intermediate-mass stars and for 
both types of magnetism, the origin of the observed magnetic field is still 
unknown \citep[e.g.][]{Lignieres14}. Various assumptions have been formulated in 
the literature fairly recently, some of which have already been more or less 
ruled out. For example, it was argued that the surface magnetic field could be 
the consequence of the emergence of buoyant concentrations of field produced by 
a powerful convective dynamo acting in the core of these stars. Unfortunately, 
these fields are not likely to be able to pop up at the stellar surface within 
the life time of the star \citep{MacGregor03}. Another hypothesis relies on the 
existence of a dynamo in the radiative envelope \citep{Spruit02, Braithwaite06b} 
but we still lack a definite answer to the question of whether such a process 
could really take place in stellar interiors \citep{Zahn07} and be efficient at 
producing the observed magnetic field. The most popular hypothesis is thus the 
one of a fossil field which would be the remnant of an early phase of the 
stellar evolution. In this case, the dichotomy between the strong dipolar fields 
of Ap stars and the sub-Gauss magnetism of the other A stars does not come up 
naturally and thus opens new questions on the origin of these stars' magnetism.
Tentative explanations for this dichotomy 
have involved the existence of an instability of a magnetic field wound-up by a 
sufficiently long-lasting differential rotation \citep{Auriere07}. Assuming a continuous initial 
distribution of fossil fields, \citet{Auriere07} proposed that below a given threshold magnetic fields would become 
unstable and that small scales would consequently be created in the horizontal direction. The observed field (i.e. the disk-averaged field)
 would in this case become very small due to cancellation effects between the opposite polarities of the 
destabilized field configuration. The unstable cases would then be responsible for the Vega-like magnetism whereas the stable cases 
(when the initial field is strong and differential rotation is rapidly quenched) would produce the strong dipolar structures of Ap stars.

The question of the angular momentum transport by magnetic fields in stellar 
radiative zone is a more general question that has also regained interest thanks 
to recent asteroseismic observations. The generality of this question arises 
from the fact that most stars possess a radiative zone somewhere in their 
interior. Initially, the problem of angular momentum transport was addressed in 
the context of the Sun's radiative core which was shown by helioseismic 
inversions to rotate almost rigidly \citep{Schou98}. Hydrodynamical transport 
processes such as meridional circulation or hydrodynamical instabilities proved 
to be inefficient at imposing such a low degree of differential rotation in this 
region \citep[see][for a review on transport processes induced by 
rotation]{Maeder00}. Evidence of processes efficient at transporting angular momentum has 
been obtained in very different types of stars such as red giant or pre-main-sequence solar-type 
stars. Indeed, 
asteroseismology applied to sub-giants and red-giants stars observed by the 
CoRoT and Kepler satellites has allowed to constrain the rotation rate 
of their core using the splitting of mixed modes \citep{Deheuvels12, 
Deheuvels14, Beck12, Mosser12a, Mosser12b}. At the end of the main 
sequence, the core of these stars contracts and thus tends to spin-up. But, when 
angular momentum transport by rotational mixing and circulation only is 
considered, the core is found to rotate two to three orders of magnitude faster 
than what is observed \citep{Eggenberger12, Marques13}. An efficient extraction 
of angular momentum has thus occurred at some point during the evolution of 
these stars, which is not reproduced by purely hydrodynamical processes. 
However, this transport process does not lead to uniform rotation since the 
asteroseismic analysis shows that a significant level of differential rotation still exists 
between the core and the envelope \citep{Deheuvels12, Deheuvels14}. A similar 
situation is encountered in pre-main sequence (PMS) solar-type stars which are 
expected to be fast rotators partly because they are contracting towards the 
main sequence. This process, together with the coupling with their circumstellar 
disk, should spin them up close to break-up velocity \citep{Hartmann89}. 
However, the measurement of rotation rates of PMS stars show that they rotate 
typically at only a tenth of their break-up velocity \citep{Bouvier86}. Clearly, 
a powerful mechanism is acting to prevent young stars from spinning up during 
their early PMS evolution. Semi-empirical models have been developed to explain the
rotational history of solar-type stars from birth to the early main-sequence.
They include the extraction of angular momentum through the interaction with a disk and magnetized 
winds while the internal transport of angular momentum is modeled by a coupling timescale between 
the radiative core and the convective envelope \citep{Gallet13}. Both an efficient transport of angular momentum 
in the radiative zone and a certain level of differential rotation are required in these models to fit the rotational 
velocities observed at the ZAMS. In all 
these cases, magneto-hydrodynamical (MHD) processes could possibly provide an additional efficient way of 
extracting angular momentum from stellar interiors and could thus explain the observed rotation 
profile in those stars.

The case of a poloidal field embedded in a differentially rotating shell is a 
canonical configuration to study the interplay between magnetic field and 
differential rotation. It has been mostly studied under the assumption of 
axisymmetry. In this case, the angular momentum evolution is governed by Ohmic 
diffusion of Alfv\'en waves propagating along the field lines. The phase mixing 
between Alfv\'en waves of adjacent field lines creates strong transverse 
gradients and controls the damping time of these waves \citep{Charbonneau92, 
Spada10}. This picture is however too simplistic as various 
non-axisymmetric instabilities are likely to develop \citep[see review 
by][]{Spruit99}, leading to a profound modification of the efficiency of the 
angular momentum transport as well as of the magnetic configuration. Under the 
conditions existing in a stellar radiative region, three main types of 
instabilities are likely to occur, which differ by the source of free energy they derive from. As reviewed by \citet{Spruit99},
gravitational energy can be released through Parker-like instabilities \citep{Parker66}, magnetic free energy through purely 
magnetic instabilities \citep[like the Tayler instability,][]{Tayler73} while shear instabilities (like the 
magneto-rotational instability) depend on the free kinetic energy in the differential rotation \citep{Balbus91}.
In the stellar context, most studies have been dedicated to the first two types of 
instabilities. The buoyancy instability has been 
particularly used to explain the formation of active regions at the surface of 
cool stars. Unstable regions of toroidal field are thought to rise through the 
convective envelope from their region of creation (the tachocline) to the 
stellar surface to create bipolar magnetic structures 
\citep{Choudhuri87,Caligari95, Emonet98,Fan08, Jouve09, Favier12, Jouve13}. The 
Parker instability has also been invoked as a possible regeneration process for 
the poloidal magnetic field in the Sun and thus thought to take a large part in 
the dynamo loop in solar-like stars \citep{Cline03}. The Tayler instability (TI) has been mostly considered to assess the stability of magnetic equilibrium solutions in stellar radiative environments. Indeed, 
\citet{Tayler73}, \citet{Wright73} and \citet{Markey73} were the first to show 
that a purely poloidal or purely toroidal magnetic field would be unstable under 
all general circumstances and that a stable equilibrium would thus have to be of 
mixed poloidal/toroidal type. This idea was then investigated more thoroughly 
including additional physical processes like rotation, and confirmed 
both by analytical studies and numerical models
\citep{Braithwaite06a,Braithwaite07,Bonanno08,Duez10, Bonanno13}. The 
magneto-rotational instability (MRI) has been extensively analyzed to understand 
the dynamics of accretion disks, in particular the transport of angular momentum 
induced by such an instability \citep[see][for a complete lecture on the 
subject]{Fromang13}. It has also been used in the context of planetary interiors 
\citep{Petitdemange08,Petitdemange13} where a modified version of the MRI was shown to be able to operate
 in the magnetospheric force balance relevant to the Earth's core.
 In stellar physics, the MRI of a purely poloidal or purely toroidal field was studied using local 
stability criteria \citep{Menou04, Masada06} or using global analysis in
spherical \citep{Arlt03} or cylindrical \citep{Ogilvie96, Rudiger14a, Rudiger14b} 
geometries, with a particular focus on the induced transport of angular momentum in the last two articles.
However, considering the level of differential rotation and the 
presence of strong magnetic fields in various situations encountered in stellar 
radiative zones, this mechanism could have attracted more attention. The 
main reason for favoring the TI over the MRI has been the strong stratification 
of stellar radiative zones, which has been thought to affect the MRI more 
severely \citep{Spruit99}.

We consider here the initial value problem of a purely poloidal field in a differentially rotating unstratified spherical shell. The origin of the magnetic field and the differential rotation is not addressed, they are assumed to be the result of physical processes at play during the star's history. The differential rotation acting on the poloidal field will naturally produce a mixed poloidal/toroidal configuration which will interact with the rotation. The initial differential rotation is not enforced by the boundary conditions, so the final state is known to be solid body rotation. In a companion paper (Gaurat et al., in prep), the evolution of the axisymmetric system was studied in 2D and tentative predictions were formulated on the stability of the induced magnetic configurations. In this work where 3D numerical simulations are performed, we wish to address the two questions quoted at the beginning of this introduction, with a particular focus on the following more specific interrogations:
\begin{itemize}
\item[-] What are the stability conditions for an initial poloidal field wound-up by differential rotation?
\item[-] What is the nature of the instability if there is any?
\item[-] What is the impact of the instability on the efficiency of the angular momentum transport?
\item[-] What is the impact of the instability on the large scale magnetic field?
\item[-] What would be the consequences for the observed magnetic fields and how does this relate to the properties of the magnetism of A-type stars?
\end{itemize}

After presenting the numerical setup in Section \ref{sect:num}, we treat separately the case of the instability of the initial purely poloidal field, without rotation or shear (Sect. \ref{sect:polo}). This enables us to show the limited impact the poloidal field instability will have on the timescales considered in the next sections. Section \ref{sect:shear} deals with the instability of the wound-up field, with various initial poloidal field configurations. In this section, a particular focus was made on the transport of angular momentum induced by the instability. We then interpret the results this work in the context of the magnetic dichotomy observed in A-type stars in Section \ref{sect:astars}. A brief summary of the consequences of this work is then given as a conclusion in Sect.\ref{sect:conclu}.


\section{Numerical model}
\label{sect:num}

\subsection{Governing Equations}

Assuming constant kinematic viscosity $\nu$, magnetic diffusivity
$\lambda$ and density $\rho_0 $, the system of equations for an incompressible magnetized fluid reads


\begin{eqnarray}
\vec{\nabla}\cdot \vec{u} &= & 0 \nonumber \\
\rho_0\lp \frac{\partial \vec{u}}{\partial t} + \vec{u}\cdot\vna\vec{u}\rp & = &
-\vna p+\frac{1}{\mu_0}\lp\vna\times \vec{B}\rp \times \vec{B} + \rho_0 \nu
\vec{\Delta} \vec{u} \nonumber \\
\frac{\partial \vec{B}}{\partial t} &=& \vna \times \lp \vec{u}\times \vec{B}\rp
+ \lambda \Delta\vec{B}.
\label{eq:syst1}
\end{eqnarray}
%
The equations are then non-dimensionalised using the viscous diffusion time $d^2/\nu$ as a reference time scale 
(where $d=r_o-r_i$ is the thickness of the spherical shell),
$\nu/d$ as a velocity scale, $\rho_0$ as a density scale and
$\sqrt{\rho_0\mu_0}\lambda/d$ as a typical scale for the magnetic field. The system
of equations (\ref{eq:syst1}) can then be rewritten as follows
  
\begin{eqnarray}
 \lp \frac{\partial \vec{u}}{\partial t} + \vec{u}\cdot\vna\vec{u}\rp &
=& -\vna p+ \frac{1}{\rm{Pm^2}}\lp\vna\times \vec{B}\rp \times \vec{B} + 
\vec{\Delta}\vec{u}, \nonumber \\
\frac{\partial \vec{B}}{\partial t} &=& \vna \times \lp \vec{u}\times \vec{B}\rp
+ \frac{1}{\rm{Pm}}\Delta\vec{B}.
\label{eq:syst2}
\end{eqnarray}
The initial seed field imposed at the beginning of the
simulation is then given in units of 

\begin{equation}
 \frac{B_0 d}{\sqrt{\rho_0\mu_0}\lambda} = \frac{t_\lambda}{t_{ap}} =
{\rm Lu_p},
\end{equation}
where $B_0$ is a measure of the surface magnetic field at 
the poles. $t_{ap} = \sqrt{\rho_0\mu_0}d/B_0$ is the Alfv\'en timescale based 
on the poloidal field and
$t_\lambda=d^2/\lambda$ is the magnetic diffusion timescale. ${\rm Lu_p} = B_0
d/\sqrt{\rho_0\mu_0}\lambda$ is the Lundquist number based on the poloidal magnetic field and
$\rm{Pm}=\nu/\lambda$ is the magnetic Prandtl number.

If, in addition to an initial seed, we also impose a differential rotation
profile, then the maximal value of the rotation rate $\Omega_0$, located on the axis of rotation, is proportional to the Reynolds number ${\rm Re}$.
The initial velocity is thus given in units of

\begin{equation}
 \frac{\Omega_0 d^2}{\nu} = {\rm Re}.
\end{equation}
This hydrodynamical problem is therefore controlled by three dimensionless
parameters, namely



\begin{equation}
 {\rm Re} = \frac{\Omega_0 d^2}{\nu} \quad;\quad {\rm Lu_p} =
\frac{B_0 d}{\sqrt{\rho_0\mu_0}\lambda} \quad;\quad
{\rm Pm}=\frac{\nu}{\lambda}.
\end{equation}
In all our calculations, the value of the magnetic Prandtl number ${\rm Pm}$ has 
been set to $1$. The study of the influence of this parameter is postponed to 
future works.


\subsection{Numerical method and boundary conditions}

The numerical simulations were computed with the code MagIC \citep{Wicht02} which solves the MHD equations in a spherical shell using a poloidal toroidal decomposition for the velocity and the magnetic fields:

\begin{equation}
{\bf u} = {\bf \nabla \times \nabla \times} \, (W \, {\bf e_r}) + {\bf \nabla \times} \, (Z \, {\bf e_r}),
\end{equation}

\begin{equation}
{\bf B} = {\bf \nabla \times \nabla \times} \, (C \, {\bf e_r}) + {\bf \nabla 
\times} \, (D \, {\bf e_r}),
\end{equation}where $W$ ($C$) and $Z$ ($D$) are the poloidal and toroidal 
potentials. The scalar potentials $W, Z, C, D$ and the pressure $p$ are 
further expanded in spherical harmonic functions up to degree $l_{max}$ in 
colatitude $\theta$ and longitude $\phi$ and in Chebyshev polynomials up to 
degree $N_r$ in the radial direction. An exhaustive description of the 
complete numerical technique can be found in \citep{Gilman81}. Typical 
numerical resolutions employed in this study range from ($N_r=65$, 
$l_{max}=170$) for the more diffusive cases (i.e. low $\rm{Lu_p}$, low 
$\rm{Re}$) to  ($N_r=97$, $l_{max}=426$) for the less diffusive ones. The spherical shell considered extends 
from longitude $\phi=0$ to $\phi=2\pi$, from colatitude $\theta=0$ to 
$\theta=\pi$ and from radius $r=r_i=\eta/(1-\eta)$ to $r=r_o=1/(1-\eta)$ where 
$\eta=r_i/r_o$ is the aspect ratio of the shell, equal to $0.3$ in all our calculations.\\

 
 For the boundary conditions, we assume stress-free boundary conditions and non-penetrating
radial velocity at top and bottom.
The magnetic field is matched to an external potential field at the top 
and the bottom boundaries:

\begin{equation}
 \vec{B} = \vna \Phi \quad \rightarrow  \quad \Delta \Phi = 0 
 \quad \rm{for} \quad r=[r_i,\ r_o].
 \label{eq:bcb}
\end{equation}

We note here that applying such potential top and bottom boundary conditions to an internal dipolar field
is not physical and thus leads to a mismatch between the interior and exterior of the domain. This mismatch 
results in the creation of a strong current sheet at the base of the domain 
which needs to be carefully treated. This is why the purely dipolar case for the initial poloidal field will not be considered here.

\subsection{Initial conditions}
\label{sect:init}

In all our calculations, we will study the stability of a magnetic field 
configuration which results from the winding up of an initial poloidal field by 
differential rotation.

For the initial poloidal field, we choose a simple configuration which has a similar structure as a pure dipole (a $l=1$, $m=0$ configuration) but with an azimuthal current (the only non-zero component) that is constant in radius. The field lines at low latitudes are closed inside the domain and open up above around $30^o$ in latitude. The advantage of this configuration is that contrary to a dipolar structure, the mismatch between the interior of the domain and the potential boundaries does not create strong current sheets likely to be unstable and to be difficult to handle numerically. The expression for the initial poloidal field is as follows:

\begin{eqnarray}
 B_r &=& 3 \cos\theta \, \frac{r-4r_o/3  + r_i^4/3r^3}{r_o (1-(r_i/r_o)^4)}, \nonumber \\
  B_\theta &=& - \frac{3\sin\theta}{2} \, \frac{3 r -8r_o/3 - r_i^4/3r^3}{r_o (1-(r_i/r_o)^4)}, \nonumber \\
  B_\phi &=& 0
\label{eq:initb}
\end{eqnarray}
This poloidal magnetic field is normalized to be equal 
to $1$ at the surface for $\theta=[0,\pi]$ .

The profile adopted for the initial differential rotation depends only on the dimensionless cylindrical radius $s=r\sin\theta$. It is normalized at $1$ on the rotation axis. It is the same profile as the one used by \citet{Arlt11an} which is consistent with a system rotating relatively fast (tending to satisfy the Taylor-Proudman theorem) and which possesses a certain degree of differential rotation, without being linearly unstable to hydrodynamical instabilities: 

 \begin{equation}
 \Omega=\frac{1}{\sqrt{1+s^4}}.
  \label{eq:inito}
 \end{equation}
Assuming such a profile, we adopt a setup where the shear parameter 
$q=\vert\partial \ln \Omega / \partial \ln s\vert \approx 1$, which constitutes an additional
assumption in our model where the shear strength is of the order of the rotation itself.
We note that when a radial (or shellular) profile is used for the differential 
rotation, a strong meridional circulation is created which quickly advects the 
poloidal field and strongly modifies its configuration before the differential 
rotation has time to build a significant toroidal field. This situation is 
not convenient since we are interested in the possibility of a wound-up field to 
become unstable. With a cylindrical profile, the amplitude of the meridional 
flow is extremely weak and the main process acting on the poloidal field will 
thus be the winding-up by the differential rotation. 

\begin{figure}[h!]
  \centering
  \includegraphics[height=7cm]{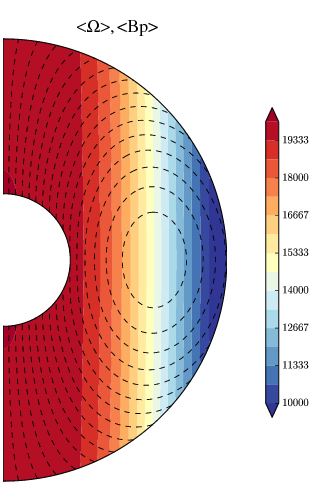}
 \caption{Azimuthal averages: initial rotation profile represented as 
isocontours of $\Omega$ (colours) and initial magnetic field represented by the 
poloidal field lines (dashed contours). $\Omega$ is given in units 
of \rm{Re}.}
 \label{fig:initb}
\end{figure}

Figure \ref{fig:initb} shows the initial velocity and magnetic field profiles used in the simulations. The magnetic field lines (dashed lines) are superimposed to the isocontours of $\Omega=v_\phi/s$ represented in colour. This type of representation helps to anticipate that the strongest winding-up of the magnetic field lines (in other words the strongest $\Omega$-effect, measured by the intensity of the term $(s \, {\bf B_p} \cdot {\bf \nabla} \Omega)$ in the induction equation) will occur where the angle between the isocontours of $\Omega$ and the magnetic field lines is maximum. From this figure, we can argue that a lobe of toroidal field will be predominantly created at mid-latitudes in each hemisphere, confined quite close to the top of the domain and that the magnetic field will change sign at the equator where the $\Omega$-effect vanishes.


\section{Instability of the initial poloidal field, without rotation and shear}
\label{sect:polo}

\subsection{Instability criterion}

In 1973, \citeauthor{Markey73} showed that, if a star contains a poloidal 
magnetic field which possesses field lines closed within the star, hydromagnetic instabilities with a growth rate 
comparable to the Alfv\'en time scale would be triggered. 
\cite{Flowers77} then showed that a poloidal field with all field lines 
crossing the stellar surface was equally subject to magnetic instabilities. In 
the case where some field lines are closed in the domain, 
the magnetic field loops 
exert pressure on each other and tend to slip out of the equilibrium position 
(or in other
words, be pushed out by pressure from the neighbours, \cite{Braithwaite07}). Since
the movement of these loops is restricted in the radial direction
by the stable stratification of the star, they move in a direction
parallel to the magnetic axis.

To look at the onset of the poloidal field instability, we start from the purely poloidal
field whose profile is given by Expression \ref{eq:initb} and shown in Fig. \ref{fig:initb} and the simulation is run without rotation or shear. Since some field lines are closed inside the domain and since a non-zero toroidal current initially exists, this configuration is a good candidate for non-axisymmetric instabilities to develop. It is expected in this case that high azimuthal wavenumbers will be excited and that the most unstable $m$ will depend on the magnetic diffusivity or, in non-dimensionalized parameters, on the Lundquist number $\rm{Lu_p}$.

\subsection{Characteristics of the instability}

The first calculation is performed for a Lundquist number $\rm{Lu_p}=200$.  
The initial perturbation is applied at all scales equivalently 
(white noise) on the poloidal magnetic field with a typical amplitude of about 
$10^{-5}$. The energy in the non-axisymmetric modes is thus initially 
around $10^{-10}$. At time $t\approx 0.8t_{ap}$, an instability 
starts to grow.

\begin{figure*}[h!]
  \centering
  \includegraphics[width=16cm]{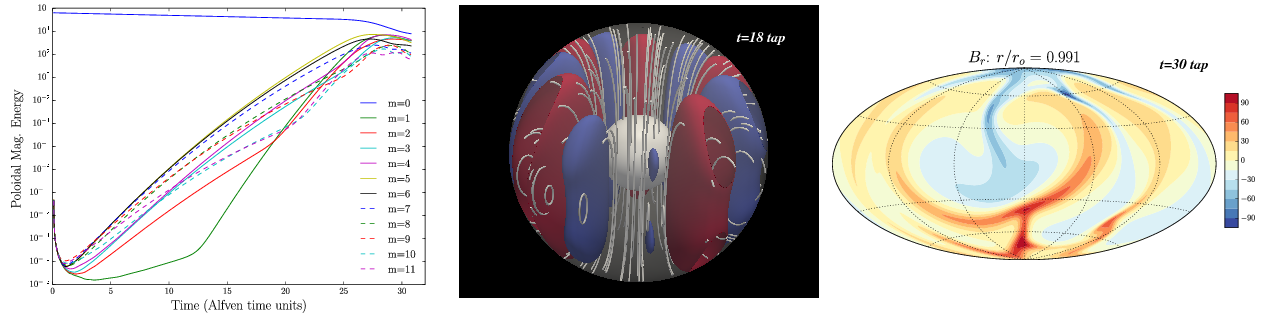}
 \caption{$\rm{Lu_p}=200$: temporal evolution of the magnetic energy contained in the first azimuthal wavenumbers (left), volume rendering showing isosurfaces of the fluctuating part of the radial magnetic field $B_r^{\prime}$ together with magnetic field lines during the linear phase of the instability (middle) and cut of the total radial magnetic field at the top of our domain when the instability has reached a saturated state.}
\label{fig:polo200}
\end{figure*}

Figure \ref{fig:polo200} shows some features of the poloidal field instability 
found in this configuration for a particular Lundquist number $\rm{Lu_p}=200$. 
The left panel shows the temporal evolution of the poloidal magnetic energy 
contained in the various azimuthal wavenumbers, from $m=0$ to 
$m=11$. In this situation, the most unstable wavenumbers are $m=5$ and $m=6$, as 
also illustrated in the mid-panel where a volume rendering of the fluctuating 
component of the radial magnetic field is shown. We find that the instability 
starts to develop close to the equator and at mid-radius, where the poloidal 
configuration possesses the strongest toroidal current. The last panel of Fig. 
\ref{fig:polo200} shows the total surface radial field when the 
instability has reached its non-linear phase. We clearly find that the initial 
simple magnetic field configuration is completely destroyed by the instability, 
leaving instead a complex non-axisymmetric field with rather small-scale 
structures very extended in latitude. This is very close to what \citet{Braithwaite06c} reported in their models of the instability of a 
poloidal field in a neutron star. They initiated their models with a different 
initial field configuration, inspired from the work of \citet{Flowers77} where 
the field is uniform in the domain and matches a potential field 
outside. They studied the instability in a compressible, stably-stratified 
atmosphere (when we are incompressible) but the typical structure of the radial 
magnetic field at the top of their domain is similar to what is found here in 
our simulations.

\begin{figure*}[h!]
  \centering
  \includegraphics[width=14cm]{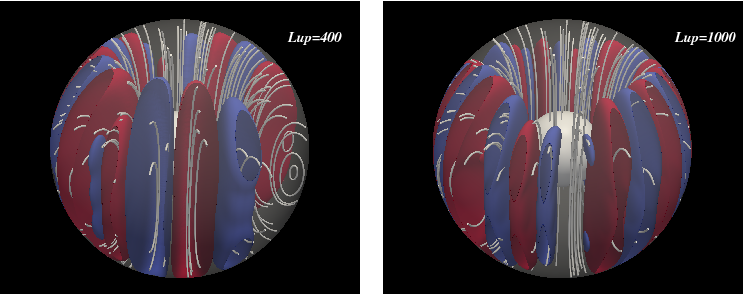}
 \caption{Volume rendering of the instability for $\rm{Lu_p}=400$ (left) and $\rm{Lu_p}=1000$ (right). Surfaces of constant $B_r^{\prime}$ are represented, as well as some magnetic field lines inside the domain.}
\label{fig:poloequat}
\end{figure*}

The effects of magnetic diffusion were then investigated by varying the 
Lundquist number $\rm{Lu_p}$. Three additional cases were calculated, with 
$\rm{Lu_p}=100$, $\rm{Lu_p}=400$ and $\rm{Lu_p}=1000$. All cases were unstable 
with respect to the poloidal field instability. The surfaces of constant 
$B_r^{\prime}$ during the linear phase of the instability are shown in 
Fig.\ref{fig:poloequat} for $\rm{Lu_p}=400$ and $\rm{Lu_p}=1000$. It is clear 
from this figure that the wavenumber with the highest growth rate (or the 
favored wavenumber) critically depends on the Lundquist 
number. Indeed, $m=4$ is favored for $\rm{Lu_p}=100$ (not shown), 
$m=10$ for $\rm{Lu_p}=400$ and $m=14-15$ for $\rm{Lu_p}=1000$, the length scale of the 
instability thus being strongly dependent on the magnetic diffusivity. The 
growth rate of the instability was measured in those various cases and it is 
found that, in agreement with theoretical studies, it is equal to the Alfv\'en 
frequency at the location of the instability, namely where the field possesses 
the strongest currents, around mid-radius and at the equator. In all cases, the 
non-linear outcome of the instability is the complete destruction of the 
large-scale axisymmetric structure into a complex non-axisymmetric field 
configuration. 

We note here that for the poloidal field instability, the unstable modes tend to 
concentrate where the current is the strongest, which is consistent 
with the fact that this instability consists in releasing the magnetic energy contained in this non-potential configuration.
 As a consequence, a purely dipolar field, which does not possess currents, should not be unstable. 
However, as stated before, imposing a dipolar field inside the domain and 
a potential field outside will necessarily create a mismatch at one of the boundaries. 
This mismatch results in a strong current sheet near the base of the domain 
which can be the seed for the development of the poloidal field instability.
This is probably what \citet{Zahn07} found when they argue that the 
high-$m$ instability of their initial dipole (which should be 
current-free and thus stable) is a classical poloidal field instability. It is 
indeed clear in their results that the instability develops at the base of their 
domain where a current sheet is artificially created because of the mismatch between their dipolar field and the 
potential field boundary conditions. We chose in this work to 
concentrate on an initial field configuration which would not lead to the 
creation of artificially strong current sheets close to the boundaries, in order 
to avoid possible numerical instabilities.   

We thus now have to take into account the fact that such a poloidal field configuration can and will become unstable on an Alfv\'en crossing time in our system. When solid-body rotation is added, \citet{Braithwaite07} showed that no significant effects were seen on the linear phase of the instability but that the non-linear outcome would be modified. In the cases we wish to study, the initial poloidal field configuration will be quickly modified by the shearing by differential rotation but our configurations may still be affected by the poloidal field instability on time scales of the order of a few poloidal Alfv\'en times. However, we will see that the instabilities that are found in a case of a wound-up field grow on a much shorter timescale and the effects of the poloidal field instability on our results will thus be very limited. 


\section{Shearing the initial poloidal field}
\label{sect:shear}

Now that we have checked the properties of the purely poloidal field instability in our simulations, we wish to study the stability of a magnetic field resulting from the action of differential rotation on this initial poloidal field. We will first analyze the large-scale axisymmetric evolution, with a particular focus on the configuration of the generated toroidal field.

\subsection{Evolution of the unperturbed system}
\label{sect:axi}

The initial evolution is a linear growth of the toroidal magnetic field 
intensity created by the winding-up of poloidal field lines by differential 
rotation (the so-called $\Omega$-effect). We quickly get two lobes of toroidal 
field of opposite sign, one in each hemisphere. This antisymmetry 
is simply due to the initial configuration of our poloidal field 
and our differential rotation which produces an $\Omega$-effect of opposite sign 
in both hemispheres. This phase is well understood by looking at the induction 
equation in an ideal MHD case in a kinematic regime (i.e. when the velocity is 
considered constant in time):
$$
\frac{\partial B_\phi}{\partial t}=s \,{\bf B_p} \cdot {\bf\nabla} \Omega
$$
The resulting toroidal field 
configuration is a relatively thin region close to the surface whose 
intensity grows in time as $B_\phi=s \,  {\bf B_p} \cdot {\bf \nabla} \Omega \, t$ as long as 
$ {\bf \nabla} \Omega$ and ${\bf B_p}$ are constant, which is approximately the case during this 
winding-up phase.

\begin{figure}[h!]
  \centering
	  \includegraphics[height=7cm]{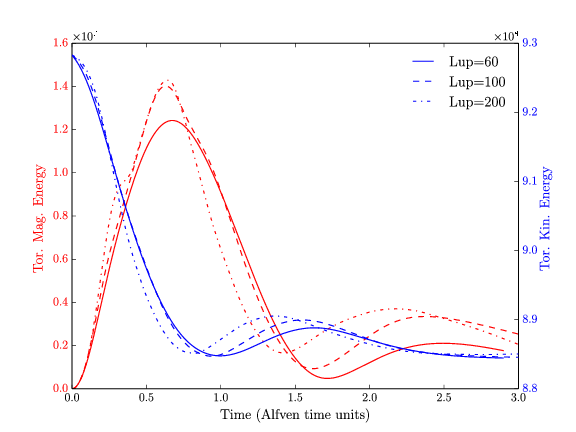}
 \caption{Temporal evolution of the toroidal magnetic (red) and kinetic (blue) energies for an initial poloidal field strength of $\rm{Lu_p}=60$ (solid lines), $\rm{Lu_p}=100$ (dashed lines) and  $\rm{Lu_p}=200$ (dash-dotted lines). In both cases, $\rm{Re}=2\times10^{4}$.}
\label{fig:axi}	
\end{figure}


An important parameter of our simulations is the ratio between the toroidal Alfv\'en frequency and the rotation rate $\omega_{A_\phi}/\Omega = B_\phi/ (r \, \Omega \sqrt{\mu_0 \rho_0})$. During the kinematic phase where the profiles of both $B_p$ and $\Omega$ are assumed to be constant, we thus expect that this ratio can be written as:
$$
\frac{\omega_{A_\phi}}{\Omega}(r,\theta,t/t_{ap})= \frac{d \,\, B_p(r,\theta,t=0) \cdot \nabla\Omega(r,\theta,t=0)}{B_0 \,\, \Omega(r,\theta,t=0)} \,\, \frac{t}{t_{ap}}
$$ 
and is independent of the poloidal Lundquist number $\rm{Lu_p}$ and on the Reynolds number $\rm{Re}$. This ratio will thus only depend on the initial and boundary conditions on the poloidal field and the rotation (see Gaurat et al., in prep.). Since the geometries of the initial poloidal field and differential rotation and the boundary conditions are kept the same in all our cases (except when the influence of the initial poloidal field configuration is investigated), this ratio will only depend on $t/t_{ap}$ in all our calculations. This property will be of prime interest for the understanding of the development of possible instabilities.
Figure \ref{fig:axi} shows the evolution of the toroidal magnetic and kinetic energies as a function of $t/t_{ap}$, for three different values of the initial poloidal field strength $\rm{Lu_p}$ and for $\rm{Re}=2\times10^{4}$. It is indeed found that in the kinematic phase, until the toroidal field reaches its maximum value, the evolution of the toroidal magnetic and kinetic energies is similar for all $\rm{Lu_p}$. The maximum value reached by the toroidal field is a bit reduced in the case where $\rm{Lu_p}=60$ since diffusive effects are not negligible here.

After this kinematic phase, the magnetic tension becomes sufficiently strong to 
be able to back-react on the differential rotation. A maximum of toroidal 
magnetic field is then reached at a time which is measured here to be 
approximately equal to $0.67 t_{ap}$ for all initial poloidal field strength 
considered. It is natural that the backreaction occurs at a time which is of the 
order of an Alfv\'en wave travel time based on the background poloidal magnetic 
field since the dynamical evolution is entirely governed by the dynamics of 
Alfv\'en waves (see Gaurat et al., in prep.).

When the toroidal magnetic field reaches its maximum, the $\Omega$-effect changes sign in the domain, thus locally creating a toroidal magnetic field of opposite sign. As seen in Fig.\ref{fig:axi}, the subsequent evolution thus consists in oscillations of $B_\phi$ and $\Omega$ with a period again approximately equal to the period of an Alfv\'en wave traveling along the poloidal field lines. Periodic oscillations are then damped and a final state of uniform rotation and zero toroidal magnetic field is reached on a time typical of magnetic diffusion acting on the length scale of Alfv\'en waves. This typical time can be rather short as small scales are created by the phase-mixing mechanism (Gaurat et al. (in prep.), \cite{Charbonneau92}). Indeed, Alfv\'en waves propagating at different speeds on neighboring surfaces progressively build-up strong gradients of field between those surfaces, thus creating a typical small scale $l$. We estimate the phase-mixing time scale to be the magnetic diffusion time based on this length scale: $t_{pm}=l^2/\lambda$. The ratio between the phase-mixing time scale and the Alfv\'en time scale $t_{ap}$ is thus given by: $t_{pm}/t_{ap}=t_{pm}/t_\lambda \times t_\lambda/t_{ap}=l^2/d^2 \times \rm{Lu_p}$. In our cases where $\rm{Lu_p} \approx 100$, we only need to produce scales 3 to 4 times smaller than the size of the domain to get a damping time of the periodic oscillations of the order of a few Alfv\'en times, as found here and as visible in Fig.\ref{fig:axi}.

\subsection{Evidence for a magnetic instability}
\label{sect:instab}

The preceding system with $\rm{Re}=2\times10^{4}$ and $\rm{Lu_p}=60$ is now 
perturbed, at an instant very close to the initial time (around 
$t=6\times10^{-3}t_{ap}$). A white noise perturbation is applied to 
the poloidal magnetic field to all combinations of $l,m$ except $l=0,m=0$ with 
the same typical amplitude $b_0^\prime=10^{-5}$. The system first continues a 
very similar evolution compared to what we get in the unperturbed case. The 
toroidal magnetic field grows linearly with time and the poloidal field and 
differential rotation stay relatively constant. However, as soon as the toroidal 
magnetic field undergoes a situation where the maximum value of 
$\omega_{A_\phi}/\Omega$ reaches the value of about $0.2$, a magnetic 
instability starts to grow.

\begin{figure*}[h!]
  \centering
	  \includegraphics[width=14cm]{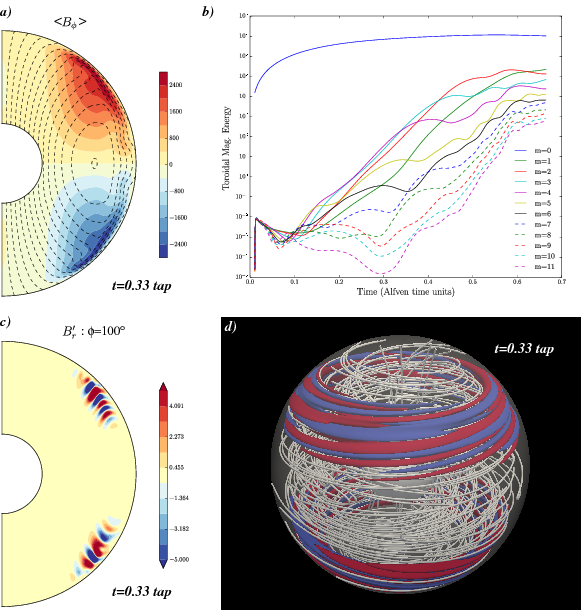}
 \caption{Characteristics of the instability for $\rm{Re}=2\times 10^{4}$ and 
$\rm{Lu_p}=60$. Panels $a)$, $c)$ and $d)$ are snapshots of respectively the 
azimuthally averaged toroidal field on which the poloidal field lines are 
superimposed, a meridional cut of  $B_r^\prime$ and a volume rendering 
showing surfaces of constant $B_r^\prime$ and magnetic field lines at $t=0.33 
t_{ap}$. Panel $b)$ shows the temporal evolution of the toroidal magnetic 
energy contained in the first 12 azimuthal wavenumbers. }
\label{fig:re2e4ha60}	
\end{figure*}

The instability is illustrated in Fig.\ref{fig:re2e4ha60}. The first panel $a$ 
shows the averaged toroidal magnetic field generated in this case in 
coloured levels, while the dashed black lines correspond to the poloidal 
field lines. Panel $b$ shows the temporal evolution of the toroidal magnetic 
energy distribution between the first 12 azimuthal wavenumbers, including the 
axisymmetric mode $m=0$. Panel $c$ shows a meridional cut of 
the fluctuating radial magnetic field (i.e. the total radial field to which the 
axisymmetric component is subtracted) and panel $d$ is a 3D rendering showing 
surfaces of constant $B_r^{\prime}$ and a few magnetic field lines. Panel $b$ 
clearly shows the exponential growth of different wavenumbers $m$, with $m=4, 
m=5$ and $m=6$ being initially favored. Around $t=0.1t_{ap}$ when the 
instability starts to grow and at the location of the development of the 
instability, $\omega_{A_\phi}/ \Omega\approx 0.2$ and $B_\phi/B_p$ is 
around $30$. We find here that the typical growth rate of the most unstable 
wavenumbers $m=4, m=5$ and $m=6$ is approximately equal to the toroidal Alfv\'en 
frequency, i.e. $\sigma/\omega_{A_\phi} \approx 1$. From panels $c$ and $d$ 
which are snapshots taken during the linear phase of the instability, we clearly 
see that this instability is confined in a region of maximum toroidal field, 
i.e. close to the surface at a latitude around $40^o$. The lengthscale of the 
instability in the latitudinal direction is rather short compared to the typical 
scale of the background magnetic field, we indeed find typically 8 nodes in the 
$\theta$-direction inside the envelope of toroidal magnetic field located 
between latitudes of $20^\circ$ and $50^\circ$.

In previous works with very similar setups \citep{Arlt11mnras,Arlt11an, 
Bonanno13}, such an instability has been attributed to the 
Tayler instability of the toroidal field. However, several arguments point to 
the fact that this is not a Tayler instability which is found here, but a 
magnetic instability of a different nature. First, the instability does not seem to be mostly located 
on the gradients of the magnetic field, which would be expected in the case of 
a Tayler instability. Indeed, let us come back to the 
criterion first derived by \citet{Goossens80} in spherical geometry for the Tayler instability of a purely toroidal field of
the form $B_\phi=f(r) P_l^1(\cos\theta)$, where $P_l^1$ is the associated Legendre polynomial of degree $l$. Since we work here with spherical coordinates, it is more directly applicable to our cases than the criterion derived by \citet{Tayler73} in cylindrical geometry. For non-axisymmetric modes (i.e. $m\neq0$), the instability occurs when:

\begin{equation}
\frac{B_\phi^2}{4\pi r^2 \sin^2\theta} \left(m^2-2\cos^2\theta -2\sin\theta \cos\theta \frac{1}{B_\phi} \frac{\partial B_\phi}{\partial \theta}\right) < 0
\end{equation}

This criterion clearly indicates that the Tayler instability 
will first develop where the latitudinal gradients of magnetic field are 
maximum. This is not what is found here. In addition, this criterion can be used to evaluate the most unstable azimuthal wavenumber.
For any configuration, the $m=1$ mode is favored but higher wavenumbers may also be excited if the gradient of $B_\phi$ is sufficiently large.
We computed this criterion for our particular configuration (even if it cannot be represented on a particular Legendre polynomial) to give us 
some hints on the possible $m$'s which could be unstable in our case. Applying Goossens' criterion, we find that the configuration is stable for $m\geq 4$.
As higher $m$'s are excited in our 3D simulation, the Tayler instability cannot solely account for the instabilities observed here.
 Finally, the growth rate of the instability does not match the 
theoretical prediction for the Tayler instability when rotation dominates (i.e. $\Omega 
>  \omega_{A_\phi}$). As stated by \citet{Szklarski13}, and as also 
found in our simulations, the maximum ratio $\omega_{A_\phi}/ \Omega$ never 
reaches values above $0.4$, and our instability starts to 
grow at $\omega_{A_\phi}/ \Omega=0.2$. We recall that, in the limit of small magnetic diffusivity,
 this maximum ratio is independent of the initial poloidal  field strength and is only related to the efficiency of the 
$\Omega$-effect which is itself only related to the initial geometry of the 
poloidal field and of the differential rotation. In cases where  $\Omega \gg 
\omega_{A_\phi}$, the TI growth rate is reduced by a factor $ \omega_{A_\phi}/ 
\Omega $ compared to the non-rotating case, where the growth rate is $ 
\omega_{A_\phi}$. This means that in our calculations, the growth rate should be 
$\sigma= \omega_{A_\phi}/ \Omega \times\,\omega_{A_\phi}= 0.2 \,\omega_{A_\phi} 
$, when we find a growth rate of $\omega_{A_\phi}$. \citet{Szklarski13} 
claim that no instability is found in any of their self-consistent 
calculations. We suspect that this is due to their stable stratification which 
probably inhibited the instability we found in our unstratified cases.

\subsection{Analysis of the instability}
\label{sect:analysis}

We argue here that this is not the Tayler instability which is found in those configurations but the magnetorotational instability of a magnetic field in the presence of a differential rotation. The configurations of magnetic field obtained in our simulations are rather complex and do not possess typical profiles for which the conditions of stability are known. In this section, we thus proceed to a local linear stability analysis, using the dispersion relation derived by \citet{Ogilvie07} for the case of a purely toroidal field with differential rotation, both possessing arbitrary profiles. The results of this analysis are then compared with the properties of the instability observed in the numerical simulation.

 In his work, \citet{Ogilvie07} develops the linear stability analysis of a 
purely toroidal magnetic field with differential rotation in an ideal MHD case. 
Instead of using a classical normal mode expansion for the disturbances, he considers a 
solution localized in the meridional plane (on the single point $s_0,z_0$) in a Gaussian envelope 
within which the displacement has a plane-wave form. 
The solutions are thus expressed as:

$$
\xi=\Re\left[\tilde{\xi}(r,\theta)\exp{(i m \phi-i\omega t)} \right]
$$

where $\tilde{\xi}$ is further expanded as:
$$
\tilde{\xi}=\hat{\xi} \exp{(i k_s s + i k_z z)} \exp\left(-\frac{(s-s_0)^2}{l^2}-\frac{(z-z_0)^2}{l^2}\right)
$$
where $\hat{\xi}$ is a constant complex vector, $l$ is the localization scale 
and ${\bf k}=k_s {\bf e_s}+k_z {\bf e_z}$ is a real wavevector. 
Instability sets in when $\Im(\omega) > 0$.

This technique was also used to show the existence of linear instabilities in the context of accretion discs \citep{Terquem96}. This analysis is relevant in the limits where the spatial scale of the instability is small compared to the scale of the envelope which is itself small compared to the typical scale of variation of the basic state:

$$
1/k \ll l \ll L
$$
where $L$ is the scale of the gradients of the background state.

 In our case here, the scale of the instability is indeed small compared to the 
envelope in which it is contained but it is less clear if the localization scale 
is small compared to the gradients of the basic state. However, we consider the 
disturbances which are found in our simulations to be sufficiently localized to 
be able to apply this linear stability analysis and gain some insight from it. 
We nevertheless keep in mind that the results may somewhat differ from 
what is found in the simulations since we are not exactly in the asymptotic 
limit of interest. Three other assumptions of this stability analysis are 
not strictly fulfilled in our simulations, namely the ideal MHD, the purely 
toroidal field, and the time-independence of the background state.

\subsubsection{Local linear stability results}

We first adapt the local dispersion relation derived by \citet{Ogilvie07} in a general case to our situation where the flow is incompressible and without gravity. We are then left with the two possible instabilities of interest: the TI and the MRI. As mentioned above, the steady and axisymmetric basic state consists in a differential rotation profile $\Omega_0 \Omega(s,z)$ where $\Omega_0$ is a measure of the 
rotation rate at the surface and a purely toroidal magnetic field $B_0 \, B(s,z)$ where $B_0$ is the maximum field strength. The dispersion relation of \citet{Ogilvie07} is non-dimensionalised using $d$ for the length scale and $B_0$ and $\Omega_0$ for the typical scales of magnetic field and rotation rate respectively. This produces a unique dimensionless parameter $\Omega_0/\omega_{A_{\phi0}}$ where the Alfv\'en frequency $\omega_{A_{\phi0}}=B_0/d\sqrt{\mu_0 \rho_0}$. It is quite noticeable here that the influence of the wavevector ${\bf k}$ will only enter
the dispersion relation through its direction, and not its norm. We thus only need to choose a particular direction for the perturbation, the dispersion relation being independent of the poloidal wavenumber itself.

If we consider the perturbation such that ${\bf k}$ is aligned with the rotation axis (i.e. a displacement in the direction ${\bf e_s}$ perpendicular to the axis since the perturbation is transverse in this analysis), we get the following form of the dimensionless dispersion relation:

\begin{eqnarray}
\left [ \omega^2-\frac{m^2 B^2}{s^2}-2 \left(\frac{\Omega_0}{\omega_{A_{\phi0}}}\right)^2 \, s \, \Omega \, {\bf e_s}\cdot{\bf \nabla}\Omega + 2 \, B \, {\bf e_s} \cdot {\bf \nabla}\left( \frac{B}{s} \right)\right ] \times \nonumber \\
\left[ \omega^2-\frac{m^2B^2}{s^2}  \right] = \left[ 2  \, \left(\frac{\Omega_0}{\omega_{A_{\phi0}}}\right) \, \omega \, \Omega + \frac{2mB^2}{s^2}  \right]^2 & &
\end{eqnarray}

where $s$ is the cylindrical radius and $m$ the azimuthal wavenumber considered.
 The ratio $\Omega_0/\omega_{A_{\phi0}}$ will be 
varied in our analysis to investigate the effect of rotation and differential rotation on the possible 
instabilities and then fixed to the value of $\Omega_0/\omega_{A_{\phi0}} \approx 
5$ found in our simulations when the instability starts to appear to compare 
with the results of the 3D calculations. The profiles of magnetic field and of 
differential rotation are arbitrary in this formulation, we can thus choose to 
focus on the profiles we get from the simulations when the instability starts to 
kick in. This dispersion relation is an algebraic quartic 
equation in $\omega$ and is thus easily solved numerically. In all cases, the 
perturbation with ${\bf k}$ parallel to the rotation axis was found to be the 
most unstable. We thus show results only for this particular direction of the 
poloidal wavevector.

The first results of this local analysis are obtained by choosing the toroidal magnetic field configuration and the differential rotation we get in the 3D simulations in the case where $\rm{Lu_p}=60$ and $\rm{Re}=2\times 10^{4}$, at approximately the time when the instability starts to grow, i.e. at $t=0.1t_{ap}$ (see Fig.\ref{fig:re2e4ha60}). At this time, the maximum ratio between toroidal and poloidal magnetic field is already equal to approximately $30$. We may thus assume that the effect of the poloidal magnetic field on the stability of the main concentration of toroidal magnetic field can be ignored in the analysis. As stated before, we also recall that at the time when the instability starts to develop in the 3D simulation, the ratio $\Omega/\omega_{A_{\phi}}$ at the maximum of toroidal field is close to the value of $5$.

\begin{figure*}[h!]
  \centering
	  \includegraphics[width=14cm]{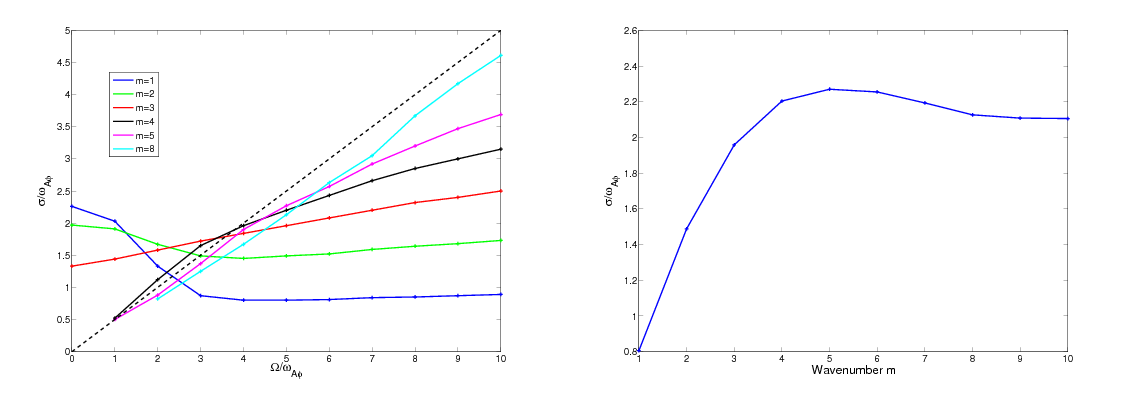}
 \caption{Maximal growth rate of the instability for the magnetic field and the differential rotation of the case with $\rm{Re}=2\times 10^{4}$ and $\rm{Lu_p}=60$ as a function of $ \Omega_0/\omega_{A_{\phi0}}$ for different values of $m$ (left) and as a function of the wavenumber $m$ for $ \Omega_0/\omega_{A_{\phi0}}=5$ (right).}
\label{fig:stab1}
\end{figure*}

Figure \ref{fig:stab1} shows the maximal growth rates of the 
magnetic instabilities in our configuration, as a function of the wavenumber 
$m$ and of the ratio $\Omega_0/\omega_{A_{\phi0}}$. The left panel represents 
the maximal growth rate (in units of the toroidal Alfv\'en frequency 
$\omega_{A_{\phi0}}$) as a function of  $\Omega_0/\omega_{A_{\phi0}}$ for 
different wavenumbers (different colours) and the right panel illustrates the 
dependence of the maximal growth rate on the azimuthal wavenumber for the 
particular value of $\Omega_0/\omega_{A_{\phi0}}=5$ obtained in the 3D 
simulation. As we can see on the left panel, our configuration is always 
subject to a magnetic instability for a certain $m$ independently of 
the value of $\Omega_0/\omega_{A_{\phi0}}$. However, the nature of 
the instability changes as this ratio is varied. Indeed, when 
$\Omega_0/\omega_{A_{\phi0}} < 1$, the dominant instability favors the $m=1$ 
mode with a growth rate $\sigma/\omega_{A_{\phi0}}\approx 2.26$. In this regime, 
the $m=2$ and $m=3$ modes are also unstable and the growth rate quickly drops to 
$0$ and below so that there is no instability for $m \geq 5$. In this regime 
where the magnetic field dominates, the main instability is the 
Tayler instability which favors the $m=1$ mode and requires strong 
gradients of field to develop. 

At the other limit where  $\Omega_0/\omega_{A_{\phi0}} > 4$, the favored mode always possesses a wavenumber equal to $\Omega_0/\omega_{A_{\phi0}}$. This is true for $m=4,5$ and $8$ but also true for the other modes not represented here. From the figure, we also find that the growth rate of this most unstable mode is approximately equal to $\Omega_0/2\omega_{A_{\phi0}}$. The oblique dashed line represented on the figure is the line of equation $y=x/2$, showing the maximum growth rate which can be reached for the favored mode. This regime is consistent with the magneto-rotational instability regime. 

Indeed, using a local linear analysis, \cite{Masada06} estimate the maximal growth rate of the non-axisymmetric MRI to be $\sigma_{max}= q\,\Omega_0/2$ for $m_{max}= \sqrt{-q^2+4q}\,\Omega_0/2 \,\omega_{A_{\phi0}}$, where we recall that the shear parameter $q$ is defined as $q=\vert\partial \ln \Omega / \partial \ln s\vert$. In our case where $q\approx 1$, we thus indeed expect a maximum growth rate of about $\sigma_{max}=\Omega_0/2$ and a $m_{max}$ close to $\Omega_0/\omega_{A_{\phi0}}$ \citep[as also found in the global analysis of][]{Ogilvie96}. To confirm that the instability found in this regime is the MRI, we artificially changed the direction of the gradient of $\Omega$ to make the rotation rate increase with the cylindrical radius $s$ instead of decrease with $s$ as we have in the simulations. In this case, the MRI disappears: all the modes are indeed stabilized for $\Omega_0/\omega_{A_{\phi0}} > 4$ .

For the intermediate regime where $1 < \Omega_0/\omega_{A_{\phi0}} < 4$, both instabilities coexist, with similar growth rates of the order of the Alfv\'en frequency. We note that in this regime, if the gradient of $\Omega$ is artificially modified to make the MRI disappear, the $m=3$ mode is stabilised, the $m=2$ mode is stabilised for $\Omega_0/\omega_{A_{\phi0}} > 2 $ and the $m=1$ mode is stabilised for $\Omega_0/\omega_{A_{\phi0}} > 2.5 $. The TI is thus visible on the $m=1$ mode for values of the rotation rate below  $\Omega_0/\omega_{A_{\phi0}} = 2.5$. Previous studies \citep[e.g.][]{Spruit99} state that in the rotating case (but without differential rotation), the growth rate of the TI is reduced by approximately a factor $\omega_{A_{\phi0}}/\Omega$ compared to the non-rotating case. This is indeed what is found here in the regime $\Omega_0/\omega_{A_{\phi0}} < 2.5$. However, for sufficiently high values of the rotation compared to the Alfv\'en speed ($\Omega_0/\omega_{A_{\phi0}} > 2.5$), the TI does not seem to just have a reduced growth rate in our differentially rotating cases, it completely disappears. We find that the whole system is then stabilised when rotation dominates in our particular setup.

The right panel of Fig.\ref{fig:stab1} shows the growth rate of the instability on our configuration, as a function of the wavenumber $m$ for  $\Omega_0/\omega_{A_{\phi0}} = 5$, i.e. in the MRI regime. 
As expected from theoretical works on the MRI and as recalled above, the growth rate increases with $m$ until it reaches a maximum value of about $q/2 \times \Omega_0$. The growth rate then seems to saturate to values close to the maximum for higher $m$'s. This indicates that several modes are supposed to  be unstable in our configuration with similar growth rates, with a maximum around $m\approx 5$. However, we note that this linear analysis does not take into account the effect of diffusion which will probably act on high wavenumbers to significantly decrease their growth rates. We thus do not expect to get very high-$m$ modes growing in our simulations.

\subsubsection{Comparison to 3D results}
\label{sect:3D}

Let us now go back to the results of the 3D calculations and make a tentative comparison with what is found in the local linear analysis. In Fig.\ref{fig:re2e4ha60}, panel $b$, the temporal evolution of each mode from $m=0$ to $m=11$ of the toroidal magnetic energy is shown. From this panel, it is rather clear that at $t\approx 0.1-0.15t_{ap}$, the most unstable modes are the $m=3$, $m=4$ and $m=5$ modes, which is consistent with what we expect at  $\Omega/\omega_{A_{\phi}} \approx 5$. We can calculate the typical growth rate of those modes and normalise it with the maximum Alfv\'en frequency, in order to compare with the previous results. We find a maximum growth rate of about $\sigma/\omega_{A_{\phi}}\approx 0.8$ in this case, as stated in section \ref{sect:instab}. When the initial poloidal field strength is increased, the same unstable modes are found since the typical value of  $\Omega/\omega_{A_{\phi}}$ is always around $5$ at instability onset. A growth rate of  $\sigma/\omega_{A_{\phi0}} \approx 1$ is then found for $\rm{Lu_p}=100$ and $\rm{Lu_p}=200$. The rotation rate was also varied in our simulations and the same instability is again found in more rapidly rotating cases. Again, for $\rm{Re}=4\times 10^{4}$ and $\rm{Lu_p}=60, 120, 200$ and $400$, the maximum growth rate is found to be of the order $\sigma/\omega_{A_{\phi}} \approx 1$ and the instability always starts to grow when  $\Omega/\omega_{A_{\phi}} \approx 5$. For all our cases where the shear parameter is fixed, we thus find a typical growth rate which only depends on the rotation rate $\Omega$, such that: $\sigma/\Omega \approx 0.2$.

\begin{figure*}[h!]
  \centering
	  \includegraphics[width=14cm]{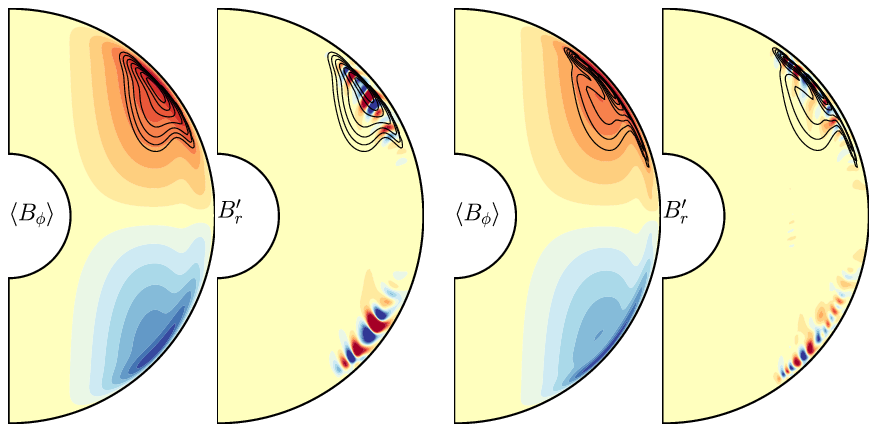}
 \caption{Azimuthally-averaged toroidal field and meridional cut of 
the fluctuating radial field during the growing phase of the instability for 
cases where  $\rm{Re}=2\times10^{4}$ and $\rm{Lu_p}=60$ (left panels) and with 
$\rm{Re}=4\times10^{4}$ and $\rm{Lu_p}=120$ (right panels). In the Northern 
hemisphere and for each snapshot, the contours of the growth rate for $m=5$ and 
$\Omega_0/\omega_{A_{\phi0}} = 5$ predicted by the linear analysis are 
superimposed to the colour contours.}
\label{fig:loc_instab}
\end{figure*}

This is in agreement with what was found in the local linear analysis 
where the maximal growth rate was dependent on the rotation rate and the shear parameter only: we had a 
maximal growth rate of $\sigma/\Omega_0 = q/2$ where $\Omega_0$ is a measure of 
the rotation rate at the surface.  
Even if the dependency of the growth rate on the rotation rate is 
recovered in our 3D calculations, we find growth rates which are typically $2$ 
to $3$ times smaller than what we expect from the local linear analysis, since 
$q\approx 1$ in our simulations. This discrepancy could be due to several 
reasons.
First, Ohmic diffusion is not taken into 
account in the linear analysis and may affect the growth rate of our instability 
which possesses rather small scales in the horizontal direction. Another reason 
could be linked to the fact that the background magnetic field on which the 
instability develops is not steady. The strength of the field, as well as its 
spatial configuration are modified during the growth of the instability. This 
evolution of the equilibrium field is of course not taken into account in the 
linear analysis and could significantly alter the instability. In particular, 
since the instability develops during the kinematic phase of the evolution of 
the axisymmetric system, the ratio  $\Omega/\omega_{A_{\phi}}$ decreases since 
the amplitude of the toroidal field grows linearly. As a consequence, the most 
unstable mode of azimuthal wavenumber $m\approx \Omega/\omega_{A_{\phi}}$ at a 
given time is not the same at a later time: the $m$ of the most unstable mode 
tends to decrease as time evolves. This is indeed what is seen in 
Fig.\ref{fig:re2e4ha60} panel $b$ where the modes of lower azimuthal 
wavenumbers seem to take over (to be of highest amplitude) as the background 
field gets amplified until the $m=2$ and then the $m=1$ modes become dominant at 
the saturation of the instability after $t\approx 0.5t_{ap}$.

Figure \ref{fig:loc_instab} shows the location of the unstable regions of 
magnetic field in a simulation with $\rm{Re}=2\times10^{4}$ and $\rm{Lu_p}=60$ 
(left panel) and with $\rm{Re}=4\times10^{4}$ and $\rm{Lu_p}=120$ (right 
panel). The azimuthally-averaged toroidal field and a meridional cut 
of the fluctuating radial field during the growing phase of the instability are 
shown, together with isocontours of the growth rate of the instability predicted 
by the linear theory for $m=5$ and $\Omega_0/\omega_{A_{\phi0}} = 5$. In both 
cases, the location of the instability in the 3D simulation agrees quite well 
with the location of the maximum growth rate predicted by the linear stability 
analysis. The instability is confined in a thin layer close to the surface 
in these cases, where the toroidal field also concentrates. We note that the 
toroidal field gets more squeezed against the top boundary in the right panel 
case where the diffusive effects are lower (the viscosity and magnetic 
diffusivity were reduced by a factor 2 compared to the case on the left panel), showing that
 some effects of diffusion are still visible on the axisymmetric evolution. 
Since the diffusion effects are decreased, smaller scales are able to exist, as 
can also be seen on the meridional structure of the fluctuating part of the 
radial field. In both cases anyway, it is the MRI which is found to be the 
dominant instability for our magnetic structure.

\subsection{Other poloidal field configurations}

Other initial poloidal fields were also tested. A classical dipolar field (with 
the bottom boundary initial mismatch) and a confined dipole were introduced with 
a similar level of magnetic energy compared to the previous case with 
$\rm{Lu_p}=100$ where the instability was clearly visible and where the effects 
of dissipation were limited. The expressions of $B_r$ and $B_\theta$ are given below for the 
dipolar field and the confined dipole (respectively Eq.\ref{eq:initbdip} and Eq.\ref{eq:initbdipconf}):

The classical formula for the dipole is used, normalized at the poles at the bottom boundary in this case:
\begin{eqnarray}
  B_r &=& \cos\theta \, \frac{r_i^3}{r^3}, \nonumber \\
  B_\theta &=& \sin\theta \,  \frac{r_i^3}{2r^3}
\label{eq:initbdip}
\end{eqnarray}

For the confined dipole, the field vanishes at the top and bottom boundaries, the field is normalized at the poles at $r=2 r_i r_o/(r_i+r_o)$:
\begin{eqnarray}
  B_r &=& 16 \cos\theta \, \frac{r_i^2 \,r_o^2 \,(r-r_i)^2 \, (r_o-r)^2} {r^4}, \nonumber \\
  B_\theta &=& 16 \sin\theta \, \frac{r_i^2 \, r_o^2 \, (r-r_i)(r_o-r)(r^2-r_i r_o)}{r^4}
\label{eq:initbdipconf}
\end{eqnarray}

\begin{figure*}[h!]
  \centering 

          \includegraphics[width=14cm]{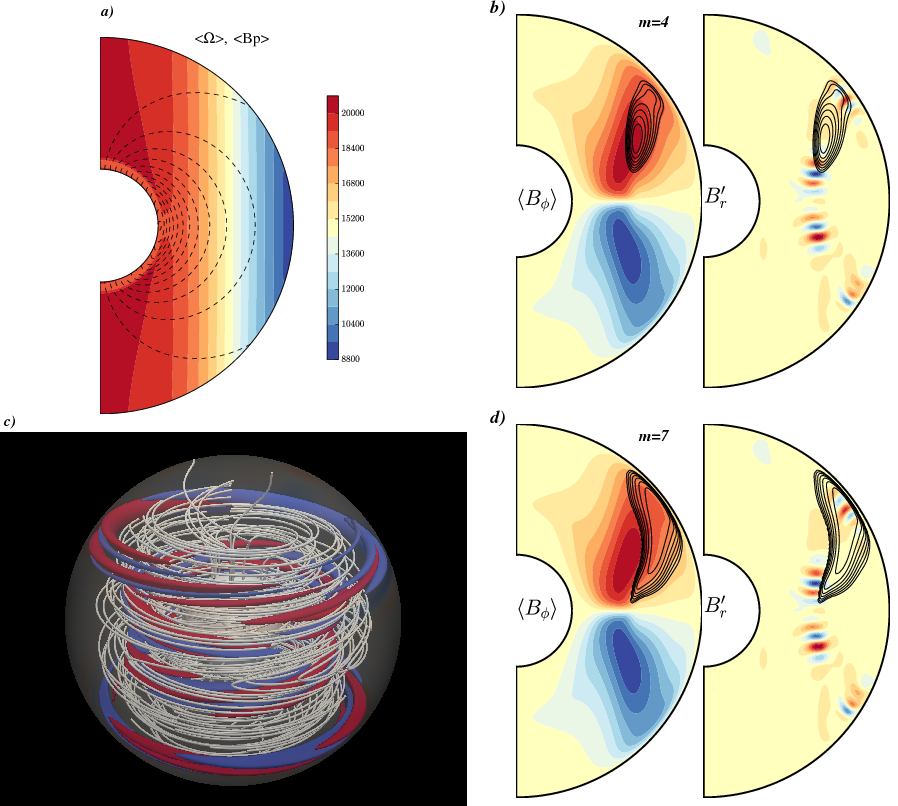}
 \caption{Instability of an initial dipole with $\rm{Re}=2\times 10^{4}$ and 
$\rm{Lu_p}=1000$ (to have the same initial poloidal energy as the reference case 
with $\rm{Lu_p}=100$). Panel $a)$: initial velocity (colours) and magnetic field 
lines (contours). Panel $c)$: volume rendering of the instability. Panels $b)$ 
and $d)$: same as Fig.\ref{fig:loc_instab} for $\Omega_0/\omega_{A_{\phi0}} = 5$ 
and $m=4$ (panel $b)$ and $m=7$ (panel $d)$. Velocity is given in units 
of \rm{Re}.}
\label{fig:dip}
\end{figure*}

\begin{figure*}[h!]
  \centering
          \includegraphics[width=14cm]{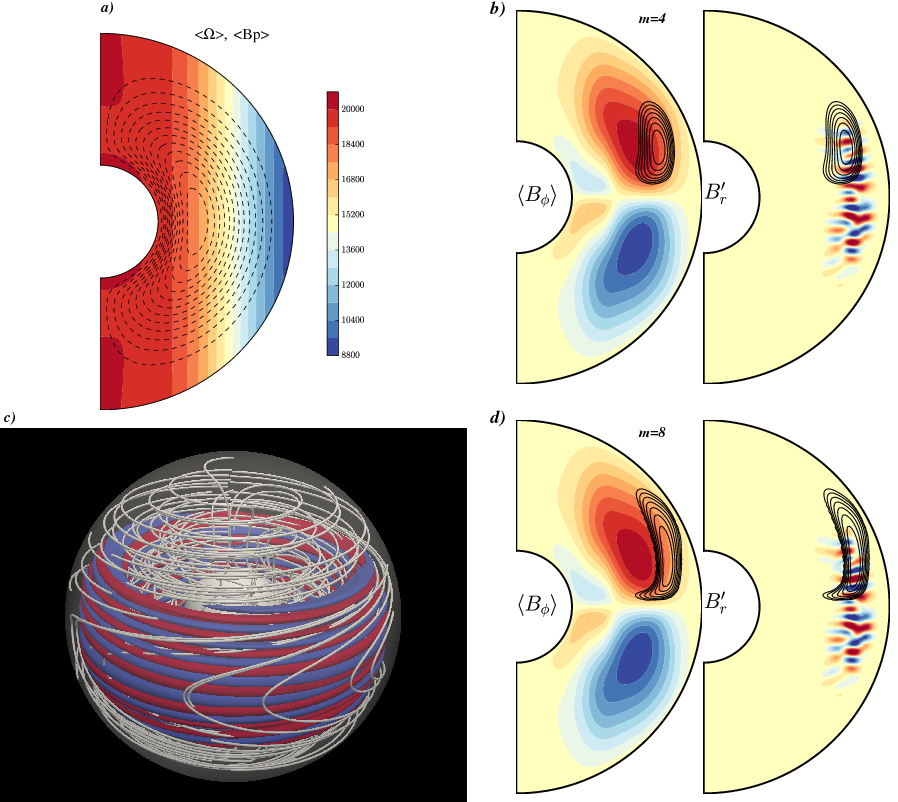}
 \caption{Instability of an initial confined dipole with $\rm{Re}=2\times 10^{4}$ and $\rm{Lu_p}=300$ (to have the same initial poloidal energy as the reference case with $\rm{Lu_p}=100$). Same as Fig.\ref{fig:dip} except that panel $d)$ shows contours of the growth rate of the $m=8$ mode.}
\label{fig:dipconf}
\end{figure*}

Similar instabilities are found in those cases but 
show different characteristics.
Similarly to the previous case, we compare the results provided by the 
local linear analysis with the characteristics of the instability found in the 
global 3D calculations. Figures \ref{fig:dip} and \ref{fig:dipconf} illustrate 
the results for both initial poloidal configurations. The shearing of the 
initial dipole by the differential rotation produces two lobes of toroidal field 
of opposite polarity, as in the previous case but much less confined in a thin 
layer close to the surface. Again, due to the initial poloidal field and 
differential rotation geometries, the maximum $\Omega$-effect is now located at 
mid-radius, close to the equator where we indeed find a maximum of toroidal 
field. This magnetic configuration is unstable to a magnetic instability which 
starts to develop at a time where $\Omega/\omega_{A_{\phi}} \approx 5$. This is 
quite remarkable that the typical value of $\Omega/\omega_{A_{\phi}}$ for which 
the instability is triggered is the same for this different initial poloidal 
field configuration. We note that this value is not reached at the same time 
$t/t_{ap}$ as with the configuration of the previous case though, since this 
time will depend on the initial field geometry, as stated in 
Sect.\ref{sect:axi}. The most unstable mode is a $m=4$ mode which 
typically grows on a toroidal Alfv\'en crossing time. Interestingly, the field 
seems to be unstable in two distinct regions, one located around the maximum of 
toroidal field inside the domain and the other one located close to the surface 
where strong radial gradients of $B_\phi$ exist, as seen on the right panels of 
Fig.\ref{fig:dip}. We note that the poloidal wavevector ${\bf k}$ tends to be 
parallel to ${\bf e_\theta}$ for the unstable modes located close to the surface 
and aligned with the rotation axis for the unstable modes of the interior. In 
this situation, the linear analysis predicts a maximum growth rate for the 
azimuthal wavenumber $m=7$ but the growth rate has almost already reached its 
maximal value of $\sigma/\omega_{A_{\phi}} \approx 2.5$ for $m=4-5$. As seen in 
Fig.\ref{fig:dip}, the location of the maximum growth rate differs significantly 
when the $m=4$ or $m=7$ modes are considered, contrary to the case with the reference poloidal 
field configuration where all unstable modes were approximately located in the same region of strong toroidal field (see Fig.\ref{fig:loc_instab}).
We indeed find here that higher azimuthal 
wavenumbers $m$ tend to be unstable close to the surface where the radial 
gradient of $B_\phi$ is strong. On the contrary, lower $m$'s concentrate inside 
the domain where the latitudinal gradient dominates. Two regions of instability 
thus also appear in the local analysis, the higher $m$ modes being 
more sensitive to the strong near-surface radial gradient of $B_\phi$.
 However, it is less clear in the full 3D calculations if the 
unstable modes located close to the surface possess higher wavenumbers. Indeed, 
the $m=4$ mode seems to dominate almost everywhere in the unstable regions.

Figure \ref{fig:dipconf} shows the results of both the 3D calculations and the 
local linear analysis for the confined dipole. Again, a magnetic instability 
typically grows in the 3D simulations on a toroidal Alfv\'en time, with 
preferred $m=1, 2$ and $3$ modes, i.e. lower than in the two previous cases. 
This can be seen on the volume rendering where the isocontours of the 
fluctuating part of the radial field are shown and look quite axisymmetric. The 
unstable modes are in this case located at low latitudes where the gradient of 
toroidal field with respect to the cylindrical radius $s$ is 
the strongest. In this case, the linear analysis predicts a growth 
rate which quickly reaches a value around  $\sigma/\omega_{A_{\phi}} \approx 
2.5$ for $m\approx 5$ but which keeps on slightly increasing for higher $m$'s. 
However, higher $m$ modes tend to be unstable on thinner and thinner regions 
around the negative gradient of $\vert B_\phi \vert$ with respect to $s$, as can 
be seen on panel $d$ of Fig. \ref{fig:dipconf} where the contours of the growth 
rate for $m=8$ are represented. These very thin regions where high 
wavenumbers are supposed to be the most unstable may have reduced growth rates 
in the 3D calculations since they may be subject to Ohmic diffusion which is not 
taken into account in the local linear analysis. Moreover, in this case, the 
instability starts to develop when the maximum ratio between toroidal and 
poloidal fields is only around $B_\phi/B_p \approx 8$. As a consequence, we may 
not be able to neglect the effect of the poloidal component on the development 
of the instability of the magnetic field. In particular, additional magnetic 
tension could act to stabilise the high-wavenumber modes and thus cause the low 
$m$ modes to be favored compared to the purely toroidal case.

In both initial configurations, we confirm that the instability is of 
magneto-rotational type, since it vanishes for a positive gradient of $\Omega$ 
with respect to $s$. As in the reference case, those configurations are also 
found to be unstable to the TI when $\omega_{A\phi}$ dominates over $\Omega$, 
which is not the case when the instability sets in in the simulation. In 
addition, the TI unstable modes are located on the regions where $\vert 
B_\phi\vert$ increases with $s$, i.e. much deeper inside the spherical shell 
than the unstable regions shown here. In the magnetic configurations studied in this work, 
we have shown that the MRI dominates over the TI. One of the key parameters which puts the configuration 
in a MRI-dominated regime is the Alfv\'en frequency to rotation rate ratio. In 
all cases shown here where an initial poloidal field is sheared by a 
cylindrical differential rotation, the induced toroidal field is always 
dominated by rotation. As a consequence, the growth rate of the MRI
 is always higher than the growth rate of the TI, 
independently of the initial field configuration. We suspect however 
that a different profile for the differential rotation, a density gradient 
through the domain or a stable stratification may impact this balance.

\subsection{Effect on the transport of angular momentum}
\label{sect:angular}

We focus in this section on the transport of angular momentum induced by the development of the instability. It is particularly interesting to focus on the effect of MHD processes on the degree of differential rotation they allow. Indeed, the stellar magnetic field has often been invoked as a good candidate for forcing a flat rotation profile in a stellar radiative envelope where other transport processes such as meridional circulation or turbulence associated with hydrodynamical instabilities were not efficient enough to reduce the differential rotation. We expect here a form of turbulence associated with the magnetic instability and we may get an efficient transport of angular momentum which we intend to measure in our 3D calculations.

\begin{figure*}[!h]
  \centering
	  \includegraphics[width=14cm]{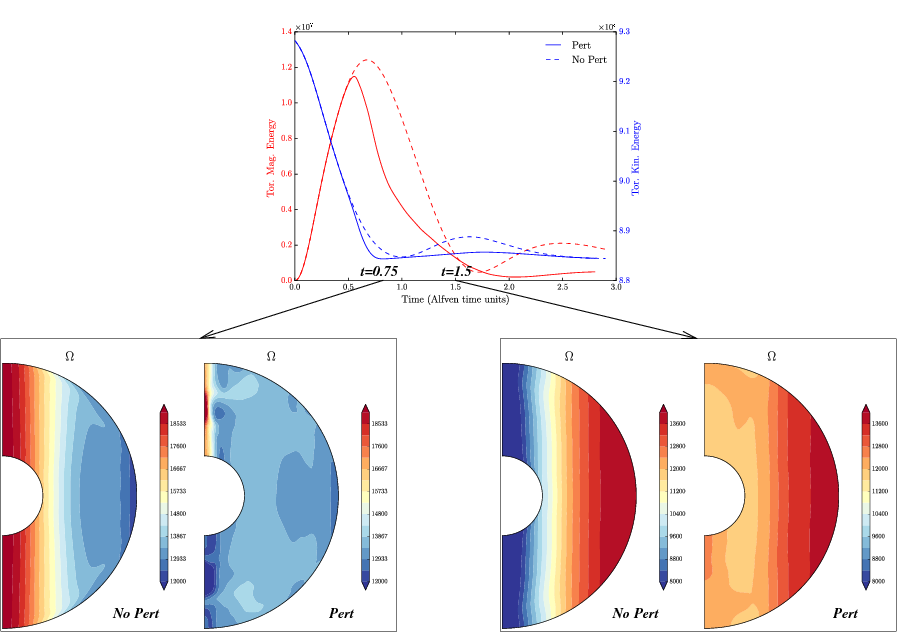}	
 \caption{Top panel: same as Fig. \ref{fig:axi} but for an unperturbed case which 
remained stable (dashed lines) and a perturbed case where the instability was 
found (solid lines). Bottom panels: azimuthally-averaged rotation rate 
for the unperturbed and perturbed cases at time $t=0.75 t_{ap}$ (left) and 
$t=1.5t_{ap}$ (right).}
\label{fig:pertVSnopert}
\end{figure*}

In Figure \ref{fig:pertVSnopert} the temporal evolution of the toroidal 
magnetic and kinetic energies are represented on the top panel, in a case with 
$\rm{Re}=2\times 10^{4}$ and $\rm{Lu_p}=60$ when a non-axisymmetric perturbation 
was applied to the system (solid lines) and without perturbation (dashed lines). 
These two cases enable to quantify the effect of the instability on the 
redistribution of angular momentum and on the evolution of the wound-up magnetic 
field, as compared to the purely axisymmetric evolution. Between $t=0$ and $t\approx 0.5 t_{ap}$, the evolution is rather similar 
between the two cases, the toroidal field takes its energy from the differential 
rotation and the magnetic energy thus increases while the kinetic energy 
decreases at the same rate. At $t\approx 0.555 t_{ap}$, a maximum of toroidal 
field is reached in the unstable case, while it is reached at $t=0.674 t_{ap}$ 
in the axisymmetric unperturbed case. As a consequence, the maximum magnetic energy is 
larger by about $8\%$ compared to the case where the instability 
developed.

 If we look at the temporal evolution of the magnetic energy in the various 
azimuthal wavenumbers $m$ of Fig.\ref{fig:re2e4ha60}, we see that 
$t=0.555 t_{ap}$ approximately corresponds to the time at which the instability 
non-linearly saturates and where the energy contained in the non-axisymmetric 
modes becomes comparable to the energy of the $m=0$ mode. The maximum value of 
the toroidal field here is thus limited by the non-linear interactions between 
modes occurring because of the development of the instability. We saw that in the 
axisymmetric case, the subsequent evolution of the magnetic field consists in 
damped oscillations with periods of the order of the Alfv\'en time. In the 
unstable case, the oscillations are also observed but their amplitude is 
significantly reduced and the final state of zero toroidal field is reached much 
faster.

Concerning the kinetic energy, a similar behaviour is found. In the unstable case, by the time the toroidal magnetic energy reaches its maximum, the kinetic energy has already almost reached its minimal value. The oscillations due to traveling Alfv \'en waves are also of much smaller amplitude than in the unperturbed case. As we said before, the final state consists of a solid body rotation. In the case where the instability developed, we can almost argue that this state of solid body rotation is reached on a time scale much closer to the Alfv\'en time scale than to the diffusive time scale, i.e. 60 times faster for this case with $\rm{Lu_p}=60$.

On the bottom panels of Fig. \ref{fig:pertVSnopert}, the averaged rotation rate 
in a meridional plane is shown for 2 different times $t=0.75 t_{ap}$ 
(left) and $t=1.5t_{ap}$ (right) in the perturbed and unperturbed cases. The 
color table is different for the two times but kept the same at each time for 
the stable and unstable cases. At time $t=0.75 t_{ap}$ when the kinetic energy 
has already reached its minimal value in the perturbed case and is still 
decreasing in the unperturbed case, the shell in the unperturbed case is still 
differentially rotating when a state of solid body rotation is already almost 
reached in the other case. For example, at the equator, the bottom boundary 
rotates $33\%$ faster than the top in the unperturbed case at $t=0.75 t_{ap}$. 
For the case where the instability acted to redistribute the angular momentum, 
this differential rotation is reduced to $0.01\%$ at the equator. Higher or 
lower values of the rotation rate still exist in this case in a thin layer close 
to the rotation axis where the toroidal field vanishes and where the 
instability was probably inefficient at transporting angular momentum since the 
unstable modes are primarily located around the strongest concentrations of 
$B_\phi$, as seen in the previous section (see Fig.\ref{fig:loc_instab}). A very 
efficient transport of angular momentum has thus occurred in the unstable case 
in the bulk of the domain, due to the strong Maxwell and Reynolds stresses 
induced by the instability. At time $t=1.5t_{ap}$ represented on the bottom 
right panels of Fig.\ref{fig:pertVSnopert}, a change of direction of the 
gradient of $\Omega$ is clearly visible in both cases but again a strong 
reduction of the differential rotation has taken place in the unstable case. The 
top boundary now rotates $36\%$ faster than the bottom in the unperturbed case 
and only $12\%$ faster in the other case. We thus find here a typical decay time 
for our differential rotation of the order of the poloidal Alfv\'en time 
$t_{decay} \approx t_{ap}$. We have yet not intended to perform a detailed 
parameter study to determine a scaling law of this decay rate with the various 
simulation parameters.

In a recent work, \citet{Rudiger14b} discusses the transport properties of an unstable toroidal magnetic field subject to the so-called Azimuthal Magneto-Rotational Instability (AMRI). For a rotation profile proportional to $1/s$ and $1/s^2$ and in cylindrical geometry, they find that the turbulent eddy viscosity $\nu_t$ associated with the AMRI scales as the rotation rate, but with a strong dependence on $\rm{Pm}$. Their value of the turbulent viscosity is in fact a maximum value since for each Reynolds number, they keep only the Lundquist number which maximizes $\nu_t$, thus leading to a scaling independent of the magnetic field strength. In their particular setup and for $\rm{Pm}=1$, they find a minimal decay time of their differential rotation of $t_{decay} \approx 200 \, t_\Omega$, the coefficient of proportionality $200$ being established numerically. The same kind of parametric study was not performed here but in typical cases, we find a decay time of the order of the poloidal Alfv\'en time $t_{decay} \approx t_{ap}$. In our notations, we have $t_{ap}=({\rm Re}/ {\rm Lu_p}) t_\Omega$ where the ratio ${\rm Re}/ {\rm Lu_p}$ was varied between $50$ to $1000$ in our simulations. We thus find a decay time which is indeed of the same order as what is found by \citet{Rudiger14b}. Similar values were also established by \citet{Arlt03} with again another setup of a purely axial field becoming unstable with respect to the MRI. In any case, this typical time to flatten the rotation profile is much faster than the magnetic diffusion time at the scale of the radius of the star. More importantly, the decay time in the unstable cases is also found to be faster than the phase-mixing time which is a diffusion time associated with the small scales created by oscillations on neighboring magnetic surfaces getting out of phase. We thus get to the conclusion that this particular instability provides an interesting additional mechanism to transport angular momentum in stellar radiative zones.


\section{Possible application to A-type stars}
\label{sect:astars}

One motivation for this type of studies is the application to the magnetic 
fields observed at the surface of intermediate-mass stars, possessing a radiative envelope. 
As said in the introduction, spectropolarimetric studies have lead to a distinction between two populations 
in those stars. The first population are the Ap stars which show a strong dipolar-like magnetic field (above $300\rm{G}$).
 The other population is formed by the non-Ap stars which do not 
exhibit strong magnetic fields. The following scenario was proposed by \cite{Auriere07} to explain this dichotomy.
If we consider an initial poloidal field and differential rotation in a radiative 
envelope, two possibilities would then occur:
\begin{itemize}
\item[-] The initial poloidal field is strong compared to the initial differential rotation, 
i.e. the winding-up time is long compared to the typical time of the back-reaction on the shear. In this 
case, the differential rotation is quickly quenched by the magnetic field and 
the configuration relaxes to a stable equilibrium. This constitutes the magnetic 
field observed in Ap stars.
\item[-] The initial poloidal field is weak compared to the initial differential rotation, 
i.e. the winding-up time is short compared to the back-reaction timescale. In this 
case, a strong toroidal field is built in a differentially rotating star. A 
magnetic instability of some type develops and creates small scales in the 
horizontal direction. Opposite polarities are created on the stellar surface and 
produce a weak line-of-sight-averaged field. The stars that underwent this 
instability would constitute the observed sub-Gauss magnetism.
\end{itemize}

\cite{Auriere07} estimated the winding-up time to be approximately $1/q\Omega$ and the
back-reaction timescale equal to $l/v_{ap}$ where $v_{ap}$ is the poloidal Alfv\'en speed and where the shear length scale $l=(\frac{\partial \ln \Omega}{\partial s})^{-1}=s/q$. The ratio 
of the two timescales, estimated at the stellar surface, will then be exactly the ratio $t_{\Omega}/t_{ap}$. The first case above thus corresponds to 
 $t_{ap} \ll t_\Omega$ and the second case to  $t_{ap} \gg t_\Omega$. At the limit  $t_{ap} = t_\Omega$, a critical field strength can be calculated: $B_{0_{crit}}= \Omega_0 d \sqrt{\rho_0\mu_0}$. For a typical main sequence Ap star (with a period of $5$ days, a radius of $3R_\odot$ and a temperature $T_{eff}=10000 \rm K$), a critical field of around $300\rm G$ is found, in agreement with the observational constraints.

\subsection{A critical number to distinguish between stable and unstable cases}


In this work, we are exactly in the right setup to test this scenario by
varying the dimensionless number ${\rm Lu_p/Re}=t_{\Omega}/t_{ap}$ in our simulations (where ${\rm Pm}=1$),
starting with the initial poloidal 
field and the cylindrical differential rotation described in 
Sect.\ref{sect:init}. We note here that the parameter ${\rm Lu_p/Re}$ measuring the ratio
of the rotation timescale $t_\Omega$ to the Alfv\'en timescale $t_{ap}$ 
was introduced in the geophysical community as the Lorentz number ${\rm Lo}$ \citep[e.g.][]{Christensen06}. 
We shall use this notation now to ease the description of our calculations.
We shall keep in mind that those calculations are 
performed assuming an incompressible fluid in an unstratified 
atmosphere, which is obviously not realistic for a stellar radiative zone. The 
same kind of study in a stably stratified spherical shell will be the subject of 
future works. However, the interaction of the magnetic field with the 
differential rotation and the development of a magnetic instability are 
processes which are well taken into account in this work. We thus wish here to 
take a first step towards testing the above scenario in a more realistic 
situation.

\begin{figure*}[h!]
  \centering
          \includegraphics[width=16cm]{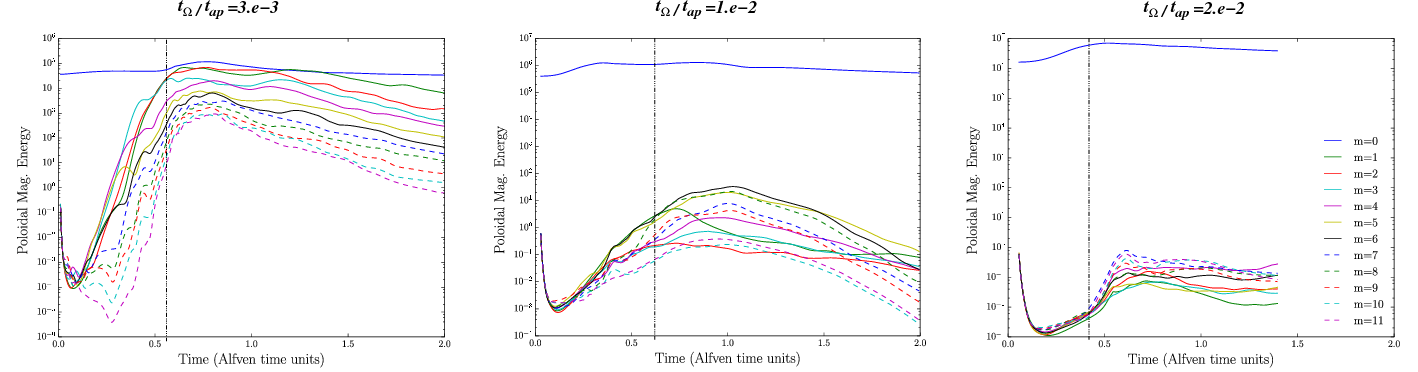}
 \caption{Temporal evolution of the poloidal magnetic energy contained in the first 12 azimuthal modes for $\rm{Re}=2\times 10^{4}$ with $\rm{Lu_p}=60$ (left), $\rm{Lu_p}=200$ (middle) and  $\rm{Lu_p}=400$ (right), corresponding to $\rm{Lo}=3\times10^{-3},10^{-2}$ and $2\times 10^{-2}$ respectively. On each snapshot, the vertical dash-dotted line represents the time at which the maximum value of the toroidal magnetic field is reached in the domain.}
\label{fig:stabVSinstab}
\end{figure*}

From Sect.~\ref{sect:3D}, the growth rate of the instability in our 
numerical simulations is equal to $\sigma = 0.2 \Omega \approx 0.1 \Omega_0$ 
since in all cases, the rotation rate at the location of the instability 
$\Omega$ is typically half the value of the rotation rate on the axis 
$\Omega_0$. This gives a typical timescale for the growth of the instability of 
$t_{inst}=1/\sigma=10/ \Omega_0=10 \, t_\Omega$. We also saw that the life time of the toroidal 
field configuration was of the order of an Alfv\'en crossing time based on the 
poloidal field $t_{ap}$. Indeed, after $t_{ap}$, regions 
of opposite polarities for the toroidal field are created due to the modified 
gradient of differential rotation. As a consequence, the instability is able to develop 
and reach the level of the axisymmetric mode only if $t_{inst}/t_{ap}= 10 \, t_\Omega/t_{ap}\ll 1$, i.e. $t_\Omega/t_{ap}=\rm{Lo} \ll 0.1$.

In Fig.\ref{fig:stabVSinstab}, the temporal evolution of the poloidal energy 
contained in the first azimuthal wavenumbers is shown for 3 different cases with 
${\rm Lo}=3\times10^{-3},10^{-2}$ and $2\times 
10^{-2}$. It is rather clear from the figure that the ratio between the dynamical time scales in our system,
has a very strong impact on the level at which the 
instability saturates. In particular, for a sufficiently small value of ${\rm Lo}$ (for example for ${\rm Lo}=3\times10^{-3}$), 
the instability has enough time to grow and 
reach the level of energy of the axisymmetric mode before the toroidal field 
gets to its maximal value. In fact, it is even the instability which saturates 
the field and prevents the toroidal field from reaching its highest possible 
value (obtained without perturbing the system, see Fig.\ref{fig:pertVSnopert}). 
On the contrary, when  ${\rm Lo}$ is increased, the 
typical growth time for the instability becomes comparable 
to the life time of the background field on which it grows. We find here that 
even when the timescale of the instability is 10 times smaller than the 
poloidal Alfv\'en crossing time ($t_\Omega/t_{ap}=10^{-2}$), 
the level of energy of the axisymmetric mode is not reached by the 
non-axisymmetric modes and the configuration will remain mostly axisymmetric. 
Similar cases were calculated with a rotation rate twice as strong and thus 
a differential rotation also twice as strong, keeping the same 
ratio between the dynamical timescales. It is also found in this 
case that $t_\Omega/t_{ap}=10^{-2}$ seems to be the critical 
value below which the axisymmetric structure will be severely altered by the 
instability and above which the system will relax into a 
quasi-axisymmetric equilibrium, which may not be that of the initial 
purely poloidal field as we will discuss in the next section.

\subsection{How does the surface magnetic field look like?}

Now that we have seen that the ratio $t_\Omega/t_{ap}$ was key to determine the level of saturation of 
the magnetic instability found in our simulations, we focus on the resulting 
field which would be observed at the stellar surface for an initial value of  
$t_\Omega/t_{ap}$ above and below the critical value quoted 
previously. We are thus here interested in the structure of the surface radial field
obtained in a particular setup where a poloidal field of arbitrary strength initially coexist 
with a cylindrical differential rotation of arbitrary strength in the stellar interior.
We expect a dramatic change in this radial field structure between cases where the instability fully develops and 
cases where the instability saturates at a low value.

\begin{figure*}[h!]
  \centering
  \includegraphics[width=14cm]{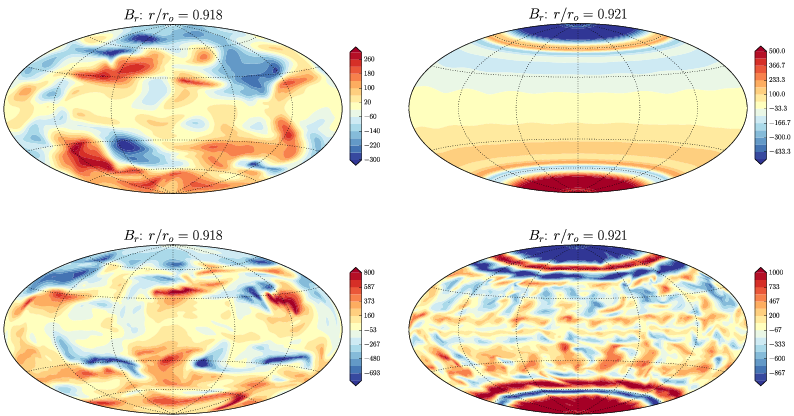}
 \caption{Radial magnetic field close to the surface at time $t=t_{ap}$ for the 
first 2 cases shown in the previous figure for ${\rm Re}=2\times 10^{4}$ (top 
panels) and for cases at ${\rm Re}=4\times 10^{4}$ (bottom panels) and with 
${\rm Lo}=3\times10^{-3}$ (left) and ${\rm Lo}=10^{-2}$ (right). 
The left panels correspond to the cases where the instability reaches the level 
of energy of the axisymmetric mode.}
\label{fig:brsurf}
\end{figure*}

Figure \ref{fig:brsurf} represents the radial component of the magnetic field 
close to the top of our computational domain for 4 different cases. The 2 left 
panels correspond to cases with ${\rm Lo}=3\times10^{-3}$ and the 2 right panels to cases with ${\rm Lo}=10^{-2}$.
 The top panels were computed with  $ {\rm Re}=2\times 
10^{4}$ and the bottom ones with ${\rm Re}=4\times 10^{4}$. This 
figure shows that the magnetic configuration dramatically differs after one 
Alfv\'en time when  ${\rm Lo}$ is increased 
from $3\times10^{-3}$ to $10^{-2}$. As expected, for a low enough value, the non-axisymmetric magnetic instability rapidly 
develops during the winding-up phase of the simulation. For both initial 
$\rm{Re}$ shown here, the resulting surface field after an Alfv\'en time is 
organized at rather small scales, strongly non-axisymmetric and the initial 
poloidal configuration has been completely modified. Since regions of opposite 
polarities exist in each hemisphere, the averaged intensity measured along the 
line-of-sight would be significantly smaller than the initial poloidal field 
strength introduced. On the contrary, when  ${\rm Lo}$ 
is increased, the MRI still grows but saturates orders of magnitude below the 
axisymmetric mode. The resulting surface field at $t=t_{ap}$ is predominantly 
axisymmetric and possesses strong polar caps reminiscent of the initial poloidal 
field considered. However, we note that this surface field does not match the 
initial poloidal field configuration. Indeed, after an Alfv\'en time, the field 
has already relaxed to a mixed toroidal/poloidal structure characterized by one 
twisted torus per hemisphere close to the poles. This can be seen in the figure 
by the belts of $B_r$ of opposite sign lying at high latitudes for both Reynolds 
number cases. This mixed poloidal/toroidal form is a classical stable 
equilibrium found both analytically \citep{Wright73} or semi-analytically 
\citep{Duez10} and numerically \citep{Braithwaite06a}.

\begin{figure}[h!]
  \centering
          \includegraphics[width=8cm]{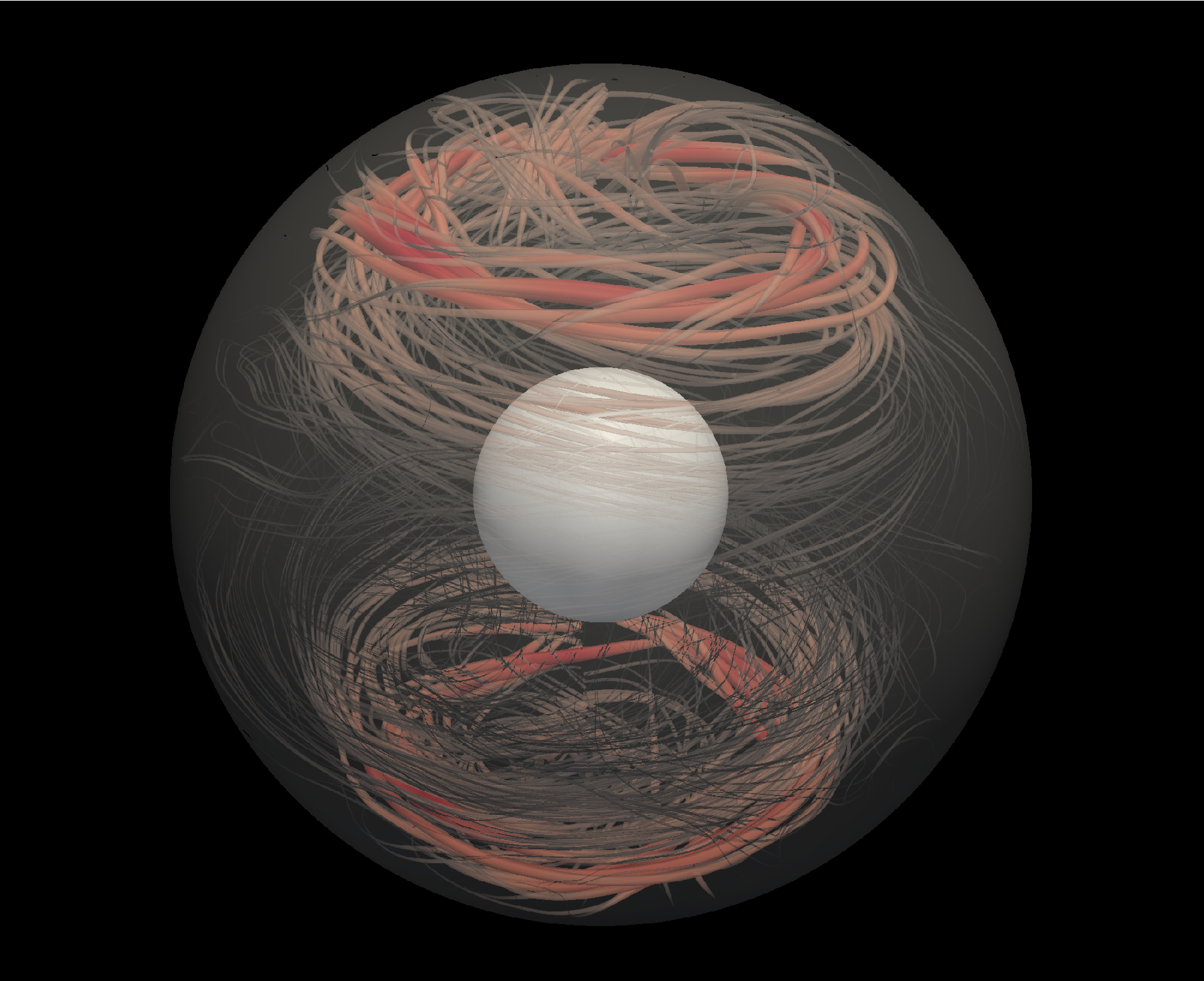}
 \caption{Magnetic field lines representation for the case at  ${\rm Re}=4\times 10^{4}$ and ${\rm Lo}=10^{-2}$ at $t=t_{ap}$. The diameter of the field lines is proportional to the magnetic energy to emphasize the structure of the magnetic field where it is maximum.}
\label{fig:torus}
\end{figure}

As stated in Sect.\ref{sect:axi}, the ratio $B_\phi/B_p$ is proportional to 
$d\Omega_0 \sqrt{\mu_0\rho_0}/B_0=t_{ap}/t_\Omega$ in the kinematic phase of the evolution, 
during which the instability grows. As a consequence, when ${\rm Lo}=t_\Omega/t_{ap}$ increases, 
the ratio between toroidal and poloidal fields 
decreases. For kink-type instabilities such as the TI, the poloidal field component stabilises
the azimuthal field and the vertical wavelength tends to be proportional to the ratio $B_p/B_\phi$ \citep[e.g.][]{Linton96}.
We could have situations (like in a stratified zone) where the vertical wavelength of possible instabilities is limited to small values (compared to the horizontal scale).
In this case, for a fixed toroidal field, the poloidal field is thus stabilising both because the field line pitch decreases (i.e. $B_p/B_\phi$ too large)
and/or because the wavelength of the possible instability is not allowed to exist in the domain. However, for the MRI found here, it is yet unclear how stabilising 
would a large poloidal field component be. If we assume that poloidal fields also have a stabilising effect on the MRI 
observed here, the higher ${\rm Lo}$ cases could also 
experiment a reduction in the growth rate of the instability because of a 
stronger impact of the poloidal field. While the ratio $B_\phi/B_p$ gets closer 
to $1$, the field thus tends to quickly relax into a mixed poloidal/toroidal 
configuration. An example of such a configuration is shown in 
Fig.\ref{fig:torus} where magnetic field lines with diameters proportional to 
the total magnetic energy are represented. This snapshot corresponds to the 
bottom right panel of Fig.\ref{fig:brsurf} where hints for a twisted 
torus close to the poles were observed. On the 3D representation of the field 
lines, it is rather clear that twisted structures have formed in both 
hemispheres, although the toroidal component is still quite dominant. As stated 
before, if the field strength of our initial poloidal field was further 
increased, the ratio between toroidal and poloidal fields would decrease and the 
twist of the field lines (and consequently their pitch angle) would increase, 
producing twisted tori with a stronger poloidal component. It is worth 
noting here that the internal magnetic field configuration in the cases where 
the instability fully developed is in fact also rather similar to one 
twisted torus per hemisphere, even if the toroidal component is stronger 
compared to the poloidal one, for the reasons just stated above. As a 
consequence, even if the observed surface radial field appears drastically 
different between the stable and unstable cases as shown in Fig. 
\ref{fig:brsurf}, the ``hidden'' internal configuration may always relax after a 
few Alfv\'en times into a mixed poloidal/toroidal configuration similar to what 
is shown in Fig.\ref{fig:torus}.   

As already stated in Sect.\ref{sect:angular}, the turbulence induced by the 
magnetic instability could produce an additional transport of angular momentum 
which causes the final state of solid-body rotation to be reached quicker in the 
perturbed cases than in the axisymmetric ones. Here, the instability develops in all cases since they are all perturbed.
But we have cases where the 
instability has fully developed to reach the level of the axisymmetric modes and 
other cases where the growth time of the instability was too long to reach a 
significant energy. By studying the differential rotation profile in each of 
these cases, we find that the angular momentum was efficiently transported again 
only in the case where the instability fully developed. A state of solid-body 
rotation is then quickly reached only when the quantity $t_\Omega/t_{ap}$ remains relatively small.  

In agreement with the scenario recalled at the beginning of this section, the ratio $t_\Omega/t_{ap}$ seems to be crucial to distinguish between cases exhibiting a large-scale axisymmetric surface magnetic field and cases where a much more complex non-axisymmetric structure dominates. A critical value of this ratio around $10^{-2}$ is found to separate the two populations. We thus find that for our initial configuration and ${\rm Pm}=1$, the critical poloidal field below which the non-axisymmetric modes dominate is:
$$
B_{0_{crit}}= 10^{-2} \Omega_0 d \sqrt{\rho_0\mu_0}
$$ 
This would indicate that the measured large-scale field in Ap stars should be at least equal to this value and that the minimum field observed in those stars should therefore be proportional to the rotation rate and to the square root of the mean density. The dependence on rotation of the minimum field in Ap stars has indeed already been observed \citep{Lignieres14} and is still being investigated through detailed spectropolarimetric surveys of additional target stars. However, we have to recall here that all the calculations were computed using a constant density, without entropy stratification and with a fixed profile for the differential rotation. The results may differ in particular on the nature of the instability and on the critical value of ${\rm Lo}$. Additional progress is needed to improve the realism of our simulations in order to compare with spectropolarimetric observations, such efforts are underway.

\section{Summary and conclusions}
\label{sect:conclu}

In this work, the interaction between differential rotation and magnetic fields 
in a simplified model of a stellar radiative zone has been investigated. 
Starting from an initial poloidal field and a cylindrical differential rotation, 
the magnetic field is first wound-up to produce a strong toroidal component 
which backreacts on the differential rotation after a time close to an Alfv\'en crossing 
time. During this winding-up phase, a non-axisymmetric magnetic instability 
develops, which may in some cases completely modify the axisymmetric structure 
and which constitutes an efficient additional transport of angular momentum. 
This instability of a wound-up field has been thought to exist and discussed in 
the literature for a long time \citep[e.g][]{Spruit99} and has recently been 
studied numerically \citep{Braithwaite06b,Arlt11an,Arlt11mnras,Szklarski13}. 
However, it has very often been interpreted as a kink-type 
Tayler instability (TI). In this work and in our particular setup, we show that the instability is in fact the 
magneto-rotational instability (MRI) of a toroidal field in the presence of a 
differential rotation. In the context of stellar physics, there were only a few studies 
\citep[e.g.][]{Arlt03, Masada06} dedicated to the MRI. In \citet{Arlt03} and \citet{Rudiger14a, Rudiger14b}, a 
particular focus on the transport of angular momentum induced by the MRI was made. They showed that 
Maxwell stresses due to magnetic field perturbations could rapidly lead to a 
state of solid-body rotation. \citet{Masada06} studied the development of magnetic 
instabilities in proto-neutron stars. They also showed that for a sufficiently 
high shear parameter, the non-axisymmetric MRI would dominate over the TI.
The crucial parameter which distinguishes the TI and the 
MRI regimes is the ratio between the rotation rate and the toroidal Alfv\'en 
frequency $\Omega/\omega_{A_{\phi}}$. In a setup where a poloidal field is 
wound-up by a cylindrical differential rotation with a shear parameter 
$q=\vert\partial \ln \Omega / \partial \ln s\vert \approx 1$ in an unstratified region,
 this ratio is always around $5$ when the instability starts to 
grow. This points towards a situation falling in the MRI regime where azimuthal 
wavenumbers of order $\Omega/\omega_{A_{\phi}}$ are the most unstable (and not 
$m=1$ like in the TI), with growth rates proportional to the rotation rate, as 
predicted by theory. The analysis of the instability is subtle since the background 
equilibrium field on which the instability develops is itself evolving with a 
characteristic timescale equal to the poloidal Alfv\'en time. As a consequence, 
the instability freely develops and strongly modifies the axisymmetric structure 
only when its growth rate (proportional to $\Omega$) is large compared to the poloidal Alfv\'en 
frequency (proportional to $B_p$) or in other words only when a weak poloidal field is 
wound-up in a rapidly rotating shell. In the opposite case, the field relaxes 
into an axisymmetric mixed poloidal/toroidal configuration. 

The results suggest that, when differential rotation is present, slowly rotating radiative envelopes possessing a strong poloidal field would tend to present a stable axisymmetric surface field whereas weaker poloidal fields embedded in more rapidly rotating zones would become unstable and create non-axisymmetric surface fields with several polarity inversions in each hemisphere. This scenario could potentially explain the magnetic dichotomy between strongly magnetic slowly rotating Ap stars and the other weakly magnetized (or non-magnetic) normal A stars. Our calculations point towards a linear dependence between the minimum field strength expected in Ap stars and their rotation rate. This relation has already been partially confirmed through detailed spectropolarimetric observations \citep{Lignieres14}. In this scenario, a source of differential rotation needs to be identified. The contraction occurring during the pre-main sequence phase might provide the required forcing to maintain a significant differential rotation. 

 We do not exclude the development of the Tayler instability (or other magnetic instabilities) for other setups which could be relevant for stellar radiative zones. In our configurations, the MRI always dominates over the TI but this may be because of the lack of stable stratification and of an overestimate of the shear parameter to be used, which is not constrained by observations. As discussed in \citet{Masada06}, a weak $N/\Omega$ ratio (where $N$ is the Brunt-V\"ais\"al\"a frequency) together with a large shear parameter $q\approx 1$ could be relevant for proto-neutron stars. In this case, the MRI could indeed be dominant. However, for parameters possibly more relevant to main-sequence stars ($N/\Omega=10^3$ and $q<0.1$), the TI would dominate for $\Omega=5\omega_{A_\phi}$. This TI regime would then probably produce a different structure of the surface magnetic field. However, in the TI regime, we still expect to recover the magnetic dichotomy: weak non-axisymmetric surface fields would result from the instability of a toroidal field-dominated configuration and strong axisymmetric stable fields would exist for poloidal and toroidal components of similar strengths.

Entropy and density stratifications have not been taken into account in this 
work and are of course needed for a correct modeling of the objects we are 
interested in. The effects of the density stratification have already been 
discussed above. As far as entropy stratification is concerned, we expect both 
instabilities to be affected by a positive value of the 
Brunt-V{\"a}is{\"a}l{\"a} frequency. It may then cause the instability found in 
our unstratified cases to be significantly reduced or limited to much 
smaller length scales although strong latitudinal gradients of differential 
rotation and magnetic fields could still be efficient at developing the MRI or 
the TI. Moreover, a stable stratification may naturally favor a 
differential rotation profile constant on concentric spheres instead of 
cylinders as was studied here \citep{Zahn92}. In the unstratified case, such a profile leads to 
a strong meridional circulation which makes the analysis of the stability of 
the wound-up toroidal field difficult. In the stratified case, a solution with 
a reduced amplitude of the meridional flow should exist, which should ease the 
study of different rotation profiles. We also note that we only studied a 
spin-up problem here, where the differential rotation was not forced by any 
external torque. The situation might be different if a forcing on the 
differential rotation was applied in the calculations, to take into account the 
presence of a convective core for example or a contraction/expansion of the 
shell. Additional numerical simulations are thus needed to efficiently tackle 
this problem and to see if non-axisymmetric instabilities can still survive and 
produce similar outcomes in more realistic situations.

\begin{acknowledgements}
T.~G. is supported by the Special Priority Program 1488 PlanetMag of the 
German Science Foundation. The authors acknowledge financial support from the Programme National
de Physique Stellaire (PNPS) of CNRS/INSU and from the Agence Nationale pour la Recherche (ANR) through the IMAGINE project. 
CPU time was provided by the HPC resources of CALMIP under the allocation 2014A-P1118. 
The authors wish to thank Fran\c{c}ois Rincon and Gordon Ogilvie for very fruitful discussions and the anonymous referee for helpful comments.
\end{acknowledgements}

\bibliographystyle{aa}
\bibliography{JGL_12_2014.bib}

\begin{thebibliography}{64}
\expandafter\ifx\csname natexlab\endcsname\relax\def\natexlab#1{#1}\fi

\bibitem[{{Arlt} {et~al.}(2003){Arlt}, {Hollerbach}, \& {R{\"u}diger}}]{Arlt03}
{Arlt}, R., {Hollerbach}, R., \& {R{\"u}diger}, G. 2003, \aap, 401, 1087

\bibitem[{{Arlt} \& {R{\"u}diger}(2011{\natexlab{a}})}]{Arlt11mnras}
{Arlt}, R. \& {R{\"u}diger}, G. 2011{\natexlab{a}}, \mnras, 412, 107

\bibitem[{{Arlt} \& {R{\"u}diger}(2011{\natexlab{b}})}]{Arlt11an}
{Arlt}, R. \& {R{\"u}diger}, G. 2011{\natexlab{b}}, Astronomische Nachrichten,
  332, 70

\bibitem[{{Auri{\`e}re} {et~al.}(2007){Auri{\`e}re}, {Wade}, {Silvester},
  {Ligni{\`e}res}, {Bagnulo}, {Bale}, {Dintrans}, {Donati}, {Folsom},
  {Gruberbauer}, {Hui Bon Hoa}, {Jeffers}, {Johnson}, {Landstreet},
  {L{\`e}bre}, {Lueftinger}, {Marsden}, {Mouillet}, {Naseri}, {Paletou},
  {Petit}, {Power}, {Rincon}, {Strasser}, \& {Toqu{\'e}}}]{Auriere07}
{Auri{\`e}re}, M., {Wade}, G.~A., {Silvester}, J., {et~al.} 2007, \aap, 475,
  1053

\bibitem[{{Balbus} \& {Hawley}(1991)}]{Balbus91}
{Balbus}, S.~A. \& {Hawley}, J.~F. 1991, \apj, 376, 214

\bibitem[{{Beck} {et~al.}(2012){Beck}, {Montalban}, {Kallinger}, {De Ridder},
  {Aerts}, {Garc{\'{\i}}a}, {Hekker}, {Dupret}, {Mosser}, {Eggenberger},
  {Stello}, {Elsworth}, {Frandsen}, {Carrier}, {Hillen}, {Gruberbauer},
  {Christensen-Dalsgaard}, {Miglio}, {Valentini}, {Bedding}, {Kjeldsen},
  {Girouard}, {Hall}, \& {Ibrahim}}]{Beck12}
{Beck}, P.~G., {Montalban}, J., {Kallinger}, T., {et~al.} 2012, \nat, 481, 55

\bibitem[{{Bonanno} \& {Urpin}(2008)}]{Bonanno08}
{Bonanno}, A. \& {Urpin}, V. 2008, \aap, 488, 1

\bibitem[{{Bonanno} \& {Urpin}(2013)}]{Bonanno13}
{Bonanno}, A. \& {Urpin}, V. 2013, \aap, 552, A91

\bibitem[{{Bouvier} {et~al.}(1986){Bouvier}, {Bertout}, {Benz}, \&
  {Mayor}}]{Bouvier86}
{Bouvier}, J., {Bertout}, C., {Benz}, W., \& {Mayor}, M. 1986, \aap, 165, 110

\bibitem[{{Braithwaite}(2006)}]{Braithwaite06b}
{Braithwaite}, J. 2006, \aap, 449, 451

\bibitem[{{Braithwaite}(2007)}]{Braithwaite07}
{Braithwaite}, J. 2007, \aap, 469, 275

\bibitem[{{Braithwaite} \& {Nordlund}(2006)}]{Braithwaite06a}
{Braithwaite}, J. \& {Nordlund}, {\AA}. 2006, \aap, 450, 1077

\bibitem[{{Braithwaite} \& {Spruit}(2006)}]{Braithwaite06c}
{Braithwaite}, J. \& {Spruit}, H.~C. 2006, \aap, 450, 1097

\bibitem[{{Caligari} {et~al.}(1995){Caligari}, {Moreno-Insertis}, \&
  {Schussler}}]{Caligari95}
{Caligari}, P., {Moreno-Insertis}, F., \& {Schussler}, M. 1995, \apj, 441, 886

\bibitem[{{Charbonneau} \& {MacGregor}(1992)}]{Charbonneau92}
{Charbonneau}, P. \& {MacGregor}, K.~B. 1992, \apj, 387, 639

\bibitem[{{Choudhuri} \& {Gilman}(1987)}]{Choudhuri87}
{Choudhuri}, A.~R. \& {Gilman}, P.~A. 1987, \apj, 316, 788

\bibitem[{{Christensen} \& {Aubert}(2006)}]{Christensen06}
{Christensen}, U.~R. \& {Aubert}, J. 2006, Geophysical Journal International,
  166, 97

\bibitem[{{Cline} {et~al.}(2003){Cline}, {Brummell}, \& {Cattaneo}}]{Cline03}
{Cline}, K.~S., {Brummell}, N.~H., \& {Cattaneo}, F. 2003, \apj, 599, 1449

\bibitem[{{Deheuvels} {et~al.}(2014){Deheuvels}, {Do{\u g}an}, {Goupil},
  {Appourchaux}, {Benomar}, {Bruntt}, {Campante}, {Casagrande}, {Ceillier},
  {Davies}, {De Cat}, {Fu}, {Garc{\'{\i}}a}, {Lobel}, {Mosser}, {Reese},
  {Regulo}, {Schou}, {Stahn}, {Thygesen}, {Yang}, {Chaplin},
  {Christensen-Dalsgaard}, {Eggenberger}, {Gizon}, {Mathis},
  {Molenda-{\.Z}akowicz}, \& {Pinsonneault}}]{Deheuvels14}
{Deheuvels}, S., {Do{\u g}an}, G., {Goupil}, M.~J., {et~al.} 2014, \aap, 564,
  A27

\bibitem[{{Deheuvels} {et~al.}(2012){Deheuvels}, {Garc{\'{\i}}a}, {Chaplin},
  {Basu}, {Antia}, {Appourchaux}, {Benomar}, {Davies}, {Elsworth}, {Gizon},
  {Goupil}, {Reese}, {Regulo}, {Schou}, {Stahn}, {Casagrande},
  {Christensen-Dalsgaard}, {Fischer}, {Hekker}, {Kjeldsen}, {Mathur}, {Mosser},
  {Pinsonneault}, {Valenti}, {Christiansen}, {Kinemuchi}, \&
  {Mullally}}]{Deheuvels12}
{Deheuvels}, S., {Garc{\'{\i}}a}, R.~A., {Chaplin}, W.~J., {et~al.} 2012, \apj,
  756, 19

\bibitem[{{Donati} \& {Landstreet}(2009)}]{Donati09}
{Donati}, J.-F. \& {Landstreet}, J.~D. 2009, \araa, 47, 333

\bibitem[{{Duez} \& {Mathis}(2010)}]{Duez10}
{Duez}, V. \& {Mathis}, S. 2010, \aap, 517, A58

\bibitem[{{Eggenberger} {et~al.}(2012){Eggenberger}, {Montalb{\'a}n}, \&
  {Miglio}}]{Eggenberger12}
{Eggenberger}, P., {Montalb{\'a}n}, J., \& {Miglio}, A. 2012, \aap, 544, L4

\bibitem[{{Emonet} \& {Moreno-Insertis}(1998)}]{Emonet98}
{Emonet}, T. \& {Moreno-Insertis}, F. 1998, \apj, 492, 804

\bibitem[{{Fan}(2008)}]{Fan08}
{Fan}, Y. 2008, \apj, 676, 680

\bibitem[{{Favier} {et~al.}(2012){Favier}, {Jouve}, {Edmunds}, {Silvers}, \&
  {Proctor}}]{Favier12}
{Favier}, B., {Jouve}, L., {Edmunds}, W., {Silvers}, L.~J., \& {Proctor},
  M.~R.~E. 2012, \mnras, 426, 3349

\bibitem[{{Flowers} \& {Ruderman}(1977)}]{Flowers77}
{Flowers}, E. \& {Ruderman}, M.~A. 1977, \apj, 215, 302

\bibitem[{{Fromang}(2013)}]{Fromang13}
{Fromang}, S. 2013, in EAS Publications Series, Vol.~62, EAS Publications
  Series, 95--142

\bibitem[{{Gallet} \& {Bouvier}(2013)}]{Gallet13}
{Gallet}, F. \& {Bouvier}, J. 2013, \aap, 556, A36

\bibitem[{{Gilman} \& {Glatzmaier}(1981)}]{Gilman81}
{Gilman}, P.~A. \& {Glatzmaier}, G.~A. 1981, \apjs, 45, 335

\bibitem[{{Goossens} \& {Tayler}(1980)}]{Goossens80}
{Goossens}, M. \& {Tayler}, R.~J. 1980, \mnras, 193, 833

\bibitem[{{Hartmann} \& {Stauffer}(1989)}]{Hartmann89}
{Hartmann}, L. \& {Stauffer}, J.~R. 1989, \aj, 97, 873

\bibitem[{{Jouve} \& {Brun}(2009)}]{Jouve09}
{Jouve}, L. \& {Brun}, A.~S. 2009, \apj, 701, 1300

\bibitem[{{Jouve} {et~al.}(2013){Jouve}, {Brun}, \& {Aulanier}}]{Jouve13}
{Jouve}, L., {Brun}, A.~S., \& {Aulanier}, G. 2013, \apj, 762, 4

\bibitem[{{Ligni{\`e}res} {et~al.}(2014){Ligni{\`e}res}, {Petit},
  {Auri{\`e}re}, {Wade}, \& {B{\"o}hm}}]{Lignieres14}
{Ligni{\`e}res}, F., {Petit}, P., {Auri{\`e}re}, M., {Wade}, G.~A., \&
  {B{\"o}hm}, T. 2014, in IAU Symposium, Vol. 302, IAU Symposium, 338--347

\bibitem[{{Ligni{\`e}res} {et~al.}(2009){Ligni{\`e}res}, {Petit}, {B{\"o}hm},
  \& {Auri{\`e}re}}]{Lignieres09}
{Ligni{\`e}res}, F., {Petit}, P., {B{\"o}hm}, T., \& {Auri{\`e}re}, M. 2009,
  \aap, 500, L41

\bibitem[{{Linton} {et~al.}(1996){Linton}, {Longcope}, \& {Fisher}}]{Linton96}
{Linton}, M.~G., {Longcope}, D.~W., \& {Fisher}, G.~H. 1996, \apj, 469, 954

\bibitem[{{MacGregor} \& {Cassinelli}(2003)}]{MacGregor03}
{MacGregor}, K.~B. \& {Cassinelli}, J.~P. 2003, \apj, 586, 480

\bibitem[{{Maeder} \& {Meynet}(2000)}]{Maeder00}
{Maeder}, A. \& {Meynet}, G. 2000, \araa, 38, 143

\bibitem[{{Markey} \& {Tayler}(1973)}]{Markey73}
{Markey}, P. \& {Tayler}, R.~J. 1973, \mnras, 163, 77

\bibitem[{{Marques} {et~al.}(2013){Marques}, {Goupil}, {Lebreton}, {Talon},
  {Palacios}, {Belkacem}, {Ouazzani}, {Mosser}, {Moya}, {Morel}, {Pichon},
  {Mathis}, {Zahn}, {Turck-Chi{\`e}ze}, \& {Nghiem}}]{Marques13}
{Marques}, J.~P., {Goupil}, M.~J., {Lebreton}, Y., {et~al.} 2013, \aap, 549,
  A74

\bibitem[{{Masada} {et~al.}(2006){Masada}, {Sano}, \& {Takabe}}]{Masada06}
{Masada}, Y., {Sano}, T., \& {Takabe}, H. 2006, \apj, 641, 447

\bibitem[{{Menou} {et~al.}(2004){Menou}, {Balbus}, \& {Spruit}}]{Menou04}
{Menou}, K., {Balbus}, S.~A., \& {Spruit}, H.~C. 2004, \apj, 607, 564

\bibitem[{{Mosser} {et~al.}(2012{\natexlab{a}}){Mosser}, {Goupil}, {Belkacem},
  {Marques}, {Beck}, {Bloemen}, {De Ridder}, {Barban}, {Deheuvels}, {Elsworth},
  {Hekker}, {Kallinger}, {Ouazzani}, {Pinsonneault}, {Samadi}, {Stello},
  {Garc{\'{\i}}a}, {Klaus}, {Li}, {Mathur}, \& {Morris}}]{Mosser12a}
{Mosser}, B., {Goupil}, M.~J., {Belkacem}, K., {et~al.} 2012{\natexlab{a}},
  \aap, 548, A10

\bibitem[{{Mosser} {et~al.}(2012{\natexlab{b}}){Mosser}, {Goupil}, {Belkacem},
  {Michel}, {Stello}, {Marques}, {Elsworth}, {Barban}, {Beck}, {Bedding}, {De
  Ridder}, {Garc{\'{\i}}a}, {Hekker}, {Kallinger}, {Samadi}, {Stumpe},
  {Barclay}, \& {Burke}}]{Mosser12b}
{Mosser}, B., {Goupil}, M.~J., {Belkacem}, K., {et~al.} 2012{\natexlab{b}},
  \aap, 540, A143

\bibitem[{{Ogilvie}(2007)}]{Ogilvie07}
{Ogilvie}, G.~I. 2007, in The Solar Tachocline, ed. D.~W. {Hughes},
  R.~{Rosner}, \& N.~O. {Weiss}, 299

\bibitem[{{Ogilvie} \& {Pringle}(1996)}]{Ogilvie96}
{Ogilvie}, G.~I. \& {Pringle}, J.~E. 1996, \mnras, 279, 152

\bibitem[{{Parker}(1966)}]{Parker66}
{Parker}, E.~N. 1966, \apj, 145, 811

\bibitem[{{Petit} {et~al.}(2011){Petit}, {Ligni{\`e}res}, {Auri{\`e}re},
  {Wade}, {Alina}, {Ballot}, {B{\"o}hm}, {Jouve}, {Oza}, {Paletou}, \&
  {Th{\'e}ado}}]{Petit11}
{Petit}, P., {Ligni{\`e}res}, F., {Auri{\`e}re}, M., {et~al.} 2011, \aap, 532,
  L13

\bibitem[{{Petitdemange} {et~al.}(2013){Petitdemange}, {Dormy}, \&
  {Balbus}}]{Petitdemange13}
{Petitdemange}, L., {Dormy}, E., \& {Balbus}, S. 2013, Physics of the Earth and
  Planetary Interiors, 223, 21

\bibitem[{{Petitdemange} {et~al.}(2008){Petitdemange}, {Dormy}, \&
  {Balbus}}]{Petitdemange08}
{Petitdemange}, L., {Dormy}, E., \& {Balbus}, S.~A. 2008, \grl, 35, 15305

\bibitem[{{R{\"u}diger} {et~al.}(2014{\natexlab{a}}){R{\"u}diger}, {Gellert},
  {Spada}, \& {Tereshin}}]{Rudiger14b}
{R{\"u}diger}, G., {Gellert}, M., {Spada}, F., \& {Tereshin}, I.
  2014{\natexlab{a}}, ArXiv e-prints

\bibitem[{{R{\"u}diger} {et~al.}(2014{\natexlab{b}}){R{\"u}diger}, {Schultz},
  \& {Kitchatinov}}]{Rudiger14a}
{R{\"u}diger}, G., {Schultz}, M., \& {Kitchatinov}, L.~L. 2014{\natexlab{b}},
  ArXiv e-prints

\bibitem[{{Schou} {et~al.}(1998){Schou}, {Antia}, {Basu}, {Bogart}, {Bush},
  {Chitre}, {Christensen-Dalsgaard}, {Di Mauro}, {Dziembowski}, {Eff-Darwich},
  {Gough}, {Haber}, {Hoeksema}, {Howe}, {Korzennik}, {Kosovichev}, {Larsen},
  {Pijpers}, {Scherrer}, {Sekii}, {Tarbell}, {Title}, {Thompson}, \&
  {Toomre}}]{Schou98}
{Schou}, J., {Antia}, H.~M., {Basu}, S., {et~al.} 1998, \apj, 505, 390

\bibitem[{{Spada} {et~al.}(2010){Spada}, {Lanzafame}, \& {Lanza}}]{Spada10}
{Spada}, F., {Lanzafame}, A.~C., \& {Lanza}, A.~F. 2010, \mnras, 404, 641

\bibitem[{{Spruit}(1999)}]{Spruit99}
{Spruit}, H.~C. 1999, \aap, 349, 189

\bibitem[{{Spruit}(2002)}]{Spruit02}
{Spruit}, H.~C. 2002, \aap, 381, 923

\bibitem[{{Szklarski} \& {Arlt}(2013)}]{Szklarski13}
{Szklarski}, J. \& {Arlt}, R. 2013, \aap, 550, A94

\bibitem[{{Tayler}(1973)}]{Tayler73}
{Tayler}, R.~J. 1973, \mnras, 161, 365

\bibitem[{{Terquem} \& {Papaloizou}(1996)}]{Terquem96}
{Terquem}, C. \& {Papaloizou}, J.~C.~B. 1996, \mnras, 279, 767

\bibitem[{{Wicht}(2002)}]{Wicht02}
{Wicht}, J. 2002, Physics of the Earth and Planetary Interiors, 132, 281

\bibitem[{{Wright}(1973)}]{Wright73}
{Wright}, G.~A.~E. 1973, \mnras, 162, 339

\bibitem[{{Zahn}(1992)}]{Zahn92}
{Zahn}, J.-P. 1992, \aap, 265, 115

\bibitem[{{Zahn} {et~al.}(2007){Zahn}, {Brun}, \& {Mathis}}]{Zahn07}
{Zahn}, J.-P., {Brun}, A.~S., \& {Mathis}, S. 2007, \aap, 474, 145

\end{thebibliography}

\end{document}